\documentclass{IEEEtran}

\usepackage{amsmath}
\usepackage{amssymb}
\usepackage{url}
\usepackage{graphicx}
\usepackage[tight]{subfigure}
\usepackage{caption}
\usepackage{multirow} 
\usepackage{rotating} 
\usepackage{float} 

\newcommand\mc{\multicolumn} 

\begin{document}

\title{A game-theoretic analysis of DoS attacks on driverless vehicles}

\date{}

\author{Ryan Shah and Shishir Nagaraja\\University of Strathclyde}
\maketitle

\begin{abstract}
Driverless vehicles are expected to form the foundation of future
connected transport infrastructure. A key weakness of connected
vehicles is their vulnerability to physical-proximity attacks such as
sensor saturation attacks. It is natural to study whether such attacks
can be used to disrupt swarms of autonomous vehicles used as part of a
large fleet providing taxi and courier delivery services. In this
paper, we start to examine the strategic options available to
attackers and defenders (autonomous-fleet operators) in such
conflicts. We find that attackers have the upper hand in most cases
and are able to carry out crippling denial-of-service attacks on
fleets, by leveraging the inherent deficiencies of road networks
identified by techniques from graph analysis.  Experimental results on
ten cities using real-world courier traces shows that most cities will
require upgraded infrastructure to defend driverless vehicles against
denial-of-service attacks. We found several hidden costs that impact
equipment designers and operators of driverless vehicles --- not
least, that road-networks need to be redesigned for robustness against
attacks thus raising some fundamental questions about the benefits.
\end{abstract}

\section{Introduction}
The area of driverless vehicles has seen rapid developments in the
last few years. Substantial industrial investment in driverless
technology has been made in the wake of recent advances in sensing and
computational control systems. While the transformative impact of such
automation has been recognised, the trust implications of their
deployment have yet to be adequately discussed.

While driverless vehicles were conceived nearly a century ago, it was
not until the application of statistical machine learning combined
with control automation, that the ideas crystalised into
reality. Therefore it is natural to ask whether it is possible to
apply adversarial statistics to also disrupt fleets of driverless
vehicles at scale. In the context of fleets, we are concerned about
the availability of driverless as the key security property, followed
by authenticity of control, integrity, and lastly confidentiality of
information. Our key concern is availability, because it is the most
easily influenced property --- an attacker who jams the optical and
acoustic channels can induce an emergency stop resulting in a
denial-of-service attack. The question however is, can this be done at
scale? Our experiments, carried out within the first systematic study
of the area, suggest that this may indeed be the case.

Driverless vehicles collect sensory input via lidars, radar,
visual-range cameras, and ultrasound sensors -- all of which are
vulnerable to signal saturation
attacks~\cite{shin:ches:2017}. Directional ultrasound acoustic jammers
consist of an array of powered ultrasound transducers whose output is
focused into a narrow beam with a distance range of a few tens of
meters~\cite{masahide:jasa:1983}. These were developed with (a short
range) for medical imaging and (medium range) for sound entertainment
systems~\cite{tseng:ijarai:2015}. Such precision engineering tools can
be repurposed as attack tools to shine an acoustic spotlight to
stealthily saturate vehicles deploying sonars. To further compromise
safety, an attacker can combine signal saturation with illusion
attacks~\cite{shin:ches:2017}. Illusion attacks cause the information
available to the car prior to jamming to be undependable thus causing
the vehicle to execute an unsafe stop.

While disabling a single driverless vehicle might not impact the
bottom line, doing so at scale certainly will. It is therefore natural
to investigate whether individual attacks can be scaled into a service
denial attack on an entire fleet of driverless vehicles. We study the
robot operator's decision-making behaviour in response to economic
costs of service-denial attacks in general (radio jamming, battery
exhaustion) with a focus on the behavioural aspects underlying
operator response. This paper is not concerned with the fulfillment of
security properties (technical jamming attacks and defences) in this
paper.

In a real-world scenario, a driverless (fully autonomous) vehicle used
to courier packages to customers would stop at the delivery address
and alert the recipient. The recipient then walks over to the vehicle,
and types in a PIN to retrieve their package. Such driverless cars are
expected to be used as part of a last-mile logistics infrastructure to
transport people (driverless taxis), deliver packages, and fresh food
at strict timelines. These are applications with trustworthiness
requirements. Indeed, online super-retailers are hoping their lastmile
problem could be solved by a fleet of driverless vehicles they
directly control, as opposed to the patchy ecosystem of multiple
providers which is the current norm.

As a more theoretical example but one that accurately captures the
threat model, consider a courier who has been tasked with delivering
bags of cocaine by a dealer. The courier sets up secret appointments
to drop off the shipments at predetermined locations. The courier must
select a route visiting all the locations and returning back to the
starting point, whilst ensuring minimum expense and maximising safety
of the goods. The courier is at risk of attacks from rival dealers who
may wish to disrupt delivery, steal goods, or worse. For easier
reading, in the rest of the paper, we refer to a driverless vehicle as
a {\em courier}.

We evaluated various  attack and defense strategies on 12 different road networks which correspond to popular cities around the world in which driverless courier vehicles are consider viable. We show that in all cases Nash equilibriums exist. However, network effects favour the attackers. Attacks can disrupt an entire fleet of driverless courier vehicles at the scale of a city  with just a few  of attack units. In some cases the defenses significantly countered an attack strategy, whereas in specific cases, the defense strategy in fact performs worse than if no strategy was employed.
 
The main contribution of this paper is to
develop the first comprehensive analysis of the attack resilience of
driverless vehicles to denial-of-service attacks on their sensory
input. Instead of analysing specific attacks we develop a generic
framework for this purpose: we consider a zero-sum multi-player
hider-seeker game where the couriers are the hiders and the attacker
coordinates one or more attack units called seekers. The couriers wish
to choose optimal routing strategies, that minimise transport time. On
the other hand, the attacker aims to choose optimal attack locations
to ambush as many couriers as possible. We consider a number of attack
and defense strategies motivated by the science of networks, and
develop a new understanding of which attack and defense strategies
result in stable equilibriums, and the implications of these findings
on the hidden costs of a robust infrastructure which is safe for
driverless vehicles.

\section{Problem description}
\label{sec:game}
Consider a courier delivery system where numerous driverless vehicles
deliver packages for purchases from an online super-retailer. Once an
online order is placed, it is added to the daily schedule of
deliveries, called a {\bf tour}, on a fleet of driverless vehicles,
individually called a {\em courier}. This can be represented using a
graph $G(V,E)$, where $G$ is the road network with vertex set $V$
corresponding to physical addresses and edgeset $E$ corresponding to
roads between them. Each courier starts from the warehouse location
$W\in V$ and ends with the warehouse location, with numerous stops at
customer delivery locations $C_1,\dots,C_l \in V$. The goal of the
couriers is to optimise both their security and transportation
time. Accordingly, each courier, modeling a {\em hider}, chooses a
routing strategy which minimises the probability of attack whilst
simultaneously minimising the delivery time (and cost). Each courier's
pure strategies are all the possible paths starting from origin
through all the destination nodes and returning back to the origin.

\begin{figure}[!h]
    \centering
    \includegraphics[width=0.4\textwidth]{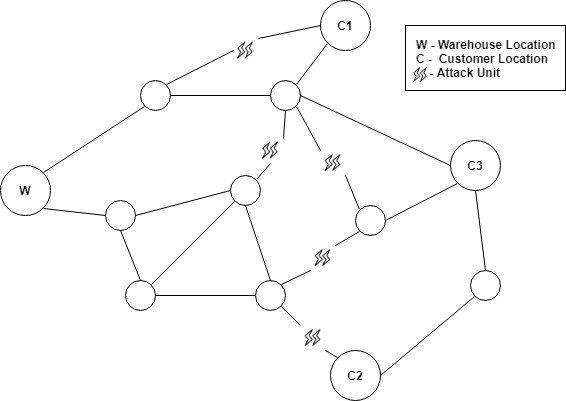}
    \caption{Multi-party Network Interdiction}
    \label{fig:ambush}
\end{figure}

The attacker's goal is to {\em ambush} as many couriers as
possible. To achieve this, the attacker positions multiple mobile
attack units each modeling a {\em seeker} on key routes (edges $S
\subset E$) of the road network in a coordinated manner, as depicted
in figure~\ref{fig:ambush}. When an attack unit comes within the
physical proximity of a driverless vehicle, it launches a
signal-saturation attacks. A signal saturation attack is a class of
targeted attacks that focuses on the sensory inputs of a courier
vehicle and orchestrates a denial-of-service by jamming the sensor. It
can take the form of targeted attacks focusing on one or more of the
following: lasers targeting the Lidar or CCD/CMOS (visual camera)
sensors, or jamming the GPS, sonar, or other sensory inputs. A
successful ambush can result in disabling the courier until it is
towed out of the attacker's proximity and possibly rebooted, hence
leading to increased delays (transportation time). The attacker's goal
is to maximise the amount of delays induced in order to maximise the
number of late deliveries. The attacker's pure strategies are all
possible combinations of attack resources to edges, thus ${_{|E|}}C_k$
strategies, where $k=|S|$.

Since the attacker has limited resources, they cannot afford to ambush
all locations all the time and instead focus on a subset of ``ideal''
locations. The set of ideal ambush locations corresponds to key routes
that are frequented by couriers that are especially useful in
achieving the courier's goals. The courier not only wants to minimise
transportation time but also ensure their robustness to ambush, thus
the courier may chose from a range of routes hoping to avoid ambush on
any of their routes.

We can model the above hider-seeker game as a multi-round zero-sum
network interdiction game~\cite{wood:mcm:1993}. For a given a road
network $G$, the couriers seek to deliver goods on time whilst hiding
from the attacker. The attacker on the other hand is an {\em
  interdictor} who aims to interdict (ambush) as many couriers as
possible.

Previous researchers~\cite{wood:mcm:1993,sanjab:corr:2017} considered
disruptive attacks on networks to be a single-round game. Such a model
is suitable for applications such as a conventional war, in which the
attacker has to expend a certain amount of effort to destroy the
defender's command, control and communications, and one wishes to
estimate how much; or a single epidemic in which a certain amount of
resource must be spent to bring the disease under control.

However, where attack and defense co-evolve in an adaptive manner,
then we have to consider a multi-round game~\cite{NA06} which has
significant explanatory power in many applications. In our scenario,
we are specifically interested in the various Nash equilibria that
might be possible with pure and mixed strategies.

Each round is consists of two phases, the attack phase and the defence
phase.  In the attack phase, the attacker deploys attack assets on a
subset of links of the network and ambushes any courier they chance
upon. The attacker selects edges according to an attack strategy
described in Section~\ref{sec:attacks}. The attacker has full
information about the roadway graph topology but has no information
about the delivery schedules of the couriers.

In the defence phase, the couriers consider the impact of the attack
on their delivery efficiency and {\em adapt} by choosing a defense
strategy in accordance with available strategy choice and
information. A defense strategy is more efficient if for a given
attack strategy, it compels the attacker to increase the number of
attack assets to achieve the same level of network disruption.

Similarly, an attack strategy is more efficient, if it either achieves
an increase in the number of successful ambushes or forces the
defender to expend more resources resulting in deliveries
beyond the delivery window.

To quantify attack efficiency, we measure the {\em percentage increase
  in late deliveries} induced by the attack as well as the increase in
the {\em delivery-completion time} -- i.e the time required to complete
deliveries per workday. We then examine the how attack efficiency
changes with variance in the delivery window size (the maximum time
that can elapse after delivery time, before a delivery is classed as
late); the impact of number of attackers on the dynamics of
attack-defense strategies.



\section{Attacks and defence strategies}
We now discuss attack and defense strategies that the players can
adopt when playing the multi-round multi-player network interdiction
game.

\subsection{Attacks}
\label{sec:attacks}

\paragraph{Random edge removal (Baseline)} The first attack strategy is the
simplest of all, and is one of the most naive attacks. A location is
chosen as a suitable site for launching service-denial attacks by
choosing an edge from the corresponding graph uniformly at
random. This models the case where an attacker has no other
information to base their choice and must choose an attack location
with no intelligence to hand.

\paragraph{Botgrep mincut detection}  Thus far, we have established that
the mincut size establishes the theoretical upper bound of
defence. Therefore it is natural to consider mincut detection
techniques as an attack strategy. The traditional description of a
mincut from a graph theory perspective, is a partition of the graph
into two disjoint subsets that are joined by a small (minimal) number
of edges.

Botgrep~\cite{nagaraja:10:botgrep} uses the relative mixing properties of
subgraphs to identify edge cuts. Botgrep uses a special probability
transition matrix to implement the random walks, where the transition
probability between adjacent nodes ${i,j} \in V$ is $min
(\frac{1}{d_i},\frac{1}{d_j})$, as opposed to $\frac{1}{d_i}$ from $i$
to $j$ in Markovian random walks, where $d_i$ is the degree of node
$i$. Botgrep uses short random walks to instrument mixing time within
a partition and to minimise the leakage of walks starting from a
partition. It then applies the probabilistic model from SybilInfer to
isolate edges which delineate the subset of the graph where mixing
speed changes. Thus the output of Botgrep is various graph subsets
with different mixing characteristics. The motivation for Botgrep is
as follows. While the notion of a {\bf small-cut} is a useful starting
point, transport networks may not necessarily contain small-cuts that
partition the graph into two or more components that are non-trivial
in size. Thus a complimentary approach to small-cut detection is
offered by the {\em Botgrep} technique which combines SybilInfer with
machine learning to identify sub-graphs with different mixing
characteristics.

\paragraph{Infomap cutset detection}
Another technique that leverages random walks is
Infomap~\cite{infomap}. The intuition underlying Infomap is that the
fraction of time spent visiting a node during a random walk can be
used to uncover dense subgraphs and the cutsets separating
them. Unlike Botgrep which uses short random walks, Infomap uses a few
long random walks to sample the graph, and computes node centralities
as a function of the number of visits during the random walk. This
information is used to search for edge cutsets partitioning the graph
using a deterministic greedy search algorithm.

\subsection{Centrality attacks}
A second class of attacks uses various measures of node centrality to
identify important nodes and proceeds to execute service denial
attacks on the node's edges. The intuition underlying these attacks is
that attackers often try to disconnect a network by destroying edges
of important (central nodes).

\paragraph{Degree centrality} The most obvious form of a node's
importance is the number of other nodes it is connected to. In this
case, the attacker targets edges of high degree nodes by deploying
attack resources on as many edges of the highest degree nodes whilst
constrained by the attack budget.

\paragraph{Eigen centrality} A related intuition of a node's importance
is not just the number of neighbouring nodes to but the importance of
the those nodes as well. A route that connects important drug routes
is even more important. Accordingly, the eigen centrality of a node is
algebraically computed as the sum of the centralities of neighbouring
nodes, which are in turn connected to many others. The highest eigen
centralities correspond to nodes located in dense
partitions. Accordingly, the edge cutset comprises edges between the
highest eigen-centrality nodes.

\paragraph{Betweenness centrality}
The principal goal of the attacker is to deploy attack resources on
edges that have the highest chance of usage. The Betweenness
centrality of a node or an edge is the fraction of the shortest paths
between all possible pairs of origin-destination pairs that include
it. Since the defender wants to make deliveries within the constraints
of distance and time, the shortest path between nodes of the tour
would be a reasonable choice for the defender although not the most
resilient choice. Accordingly, the attacker targets the set of edges
with the highest Betweenness centrality in the graph. High centrality
edges usually form a cutset separating dense subgraphs.

\subsection{Modularity attacks}
\label{sec:attackslast}
An alternate approach to mincut detection is offered by the notion of
{\em modularity}. Modularity techniques uncover mincuts that partition
a graph into two or more modules. The intuition behind modularity
mincut detection is to search for graph components which have less
edges than {\em expected} from an equivalent baseline.  The baseline is a
random graph~\cite{ER59} where the expected probability of an edge
$(i,j)$ is $d_id_j/|E|$. Modularity of graph $G(V,E)$ is accordingly defined as:
\[ Q = \sum_{i,j} A_{ij} - \frac{d_id_j}{|E|} \] where $A$ is the corresponding
adjacency matrix of graph $G$. To find the modularity mincut, a search
algorithm (for modularity optimisation) is used to detect edgesets
that can partition the graph into maximally modular subgraphs. A
number of optimisation approaches have been proposed.

\paragraph{Greedy-modularity} Clauset, Newman, and
Moore~\cite{alg:fastgreedymodulation} propose a greedy algorithm to
detect modularity mincuts. Starting from a set of disconnected nodes,
the edges of the original graph are iteratively added in order to
produce the largest possible increase of the modularity at each
step. It has a complexity of $O(N Log^2 N)$ on sparse topologies such
as road networks.

\paragraph{Eigen-modularity} Newman~\cite{N06} proposed a spectral
optimisation approach to modularity maximisation. It combines the
intuitions of eigen centrality (node centrality is recursively defined
in terms of the centrality of its neighbours) with modularity
(expected vs actual edge probability distributions) to isolate
mincuts. This method works by calculating the most significant
eigenvector of the modularity matrix defined as $B_{ij} = A_{ij} -
d_id_j/|E| $, where the first term is the adjacency matrix and the
second term is the expected edge probability according to a randomised
baseline. The graph is split into two partitions based on the sign of
the corresponding element in the eigenvector, with the mincut being
the set of edges across the two partitions. When there is no
underlying structure to leverage, the eigenvector elements are of the
same sign with the method returning a null cutset ( as opposed to
partitioning the graph into two partitions regardless of underlying
structure).

\paragraph{Hierarchical-modularity}
Blondel et. al~\cite{louvain}
propose a hierarchical modularity optimisation technique for mincut
detection. It starts with a set of isolated nodes, each within its own
partition. Edges are added from the original graph in order to produce
the maximum possible increase in modularity. In each iteration, edges
and nodes may be reassigned for merging with a different partition
with which it achieves the highest contribution to modularity. Each
partition is replaced replaced by supernodes, yielding a smaller
weighted network.  The process is then iterated, until modularity
(which is always computed with respect to the original graph) does not
increase any further.  This method offers a fair compromise between
the accuracy of the estimate of the modularity maximum, which is
better than that delivered by greedy techniques like the one by
Clauset et. al~\cite{alg:fastgreedymodulation}, and computational
complexity, which is essentially linear in the number of links of the
graph.

\subsection{Defense Strategies}
\label{sec:defenses}
\subsubsection{Naive defenses}
Our first defense strategy is the simplest of all, and is one that has
been proposed by past work in network interdiction
game~\cite{wood:mcm:1993,sanjab:corr:2017}. The defender navigates via
a randomly chosen route to reach destinations on a tour. This is
equivalent to the defender undertaking a Markovian random walk to
complete deliveries on time. The intuition behind this defense is that
by making random choices about the next part of the route the defender
can hope to maximise the attacker's uncertainty about the defender's
current location. Sanjab et. al~\cite{sanjab:corr:2017} also provide a
proof of the optimality of this defense.

The alternate obvious defense, is to enumerate all the disjoint paths
between source-destination pairs and simply choose one of the routes
uniformly at random. One might hope that there is enough redundancy
within the network structure that multiple (disjoint) path routes
exist and that the defender gets through most of the times and absorbs
the delays on account of any attacks (being towed out of the attack
zone).

Nice as these ideas may seem in theory, we find they do not work at
all well in practice. In the Evaluation section,
(Section~\ref{sec:results}) we examine the effectiveness of naive
defenses against all the attack strategies and show that they are
mostly ineffective when examined against real datasets.

\subsubsection{Sophisticated defenses}

Better results can be expected by designing defenses that are
independent of attack strategies, inspired by a common heuristic for
solving zero-sum games. The idea behind this approach is that the
defender is indifferent to attack moves, resulting in a lower-bound of
defence utility.

\paragraph{Inverse centrality defence}
Accordingly, our first non-naive {\em inverse centrality} defense is
a simple route-finding strategy that avoids edges that are commonly
used as part of a shortest-path route between origin-destination
pairs. And, where the attacker might hope to achieve a high rate of
ambush. Specifically, the defender chooses routes to explicitly
avoid high-centrality edges. The defender scores each edge $i,j$ as a
combination of Degree, Betweenness, and Eigen-centrality over node
$j$. Degree centrality is easily computed as a local metric, and it is
the most approximate form of a node's significance $C^D_{ij}=D_j/|E|$.
Betweenness centrality is roughly the proportion of paths a node lies
on $C^B_{ij}=\sum_{s \neq j \neq t} \rho_{st}(j)/\rho_{st}$, where
$\rho_{st}$ is the total number of paths between $s$ and $t$ while
$\rho_{st}(j)$ is the number of $st$ paths that include node $j$. And,
eigen centrality further incorporates the notion of node significance
as a function of being connected to significant nodes
$C^E_{ij}=C^E_{j}=\frac{1}{\lambda}\sum_{v \in V} A_{jv}C^E_{v}$ where
$\lambda$ is the largest eigenvalue. A defender needs to balance
between the various types of centrality hence they compute the
harmonic mean over the three measures as follows:

\( C_{ij} = \frac{C^D_{ij} C^B_{ij} C^E_{ij}}{C^D_{ij} + C^B_{ij} + C^E_{ij}}\)

The defender then selects a route $p$ that minimises the cumulative
weight of the scores within the path i.e \( D^p = \min_{e \in p} C_{e}
\).

\paragraph{Mixnet routing defence}
Our next defense is more sophisticated and derives from the theory of
anonymous communications as developed in traffic analysis
literature~\cite{D03}. The defender assigns a random score (between
$0$ and $1$) to each edge in the graph, and computes a route that
minimises the cumulative weight of the route whilst completing the
tour. This model was initially presented in Danezis' work on routing
anonymously in sparse networks cited above, so we refer to it as
Mixnet-shortest-path routing. In Danezis' model, each mix-router
routes messages by forwarding them to a random neighbouring mix-router
in order to maximise a global-passive adversary's uncertainty about a
message's location within a mix network. This is strikingly similar to
the defender's interest in resisting an attacker in our scenario.
The Mixnet defense promises significant improvement over Markov chain
routing proposed by previous work~\cite{sanjab:corr:2017}. Mixnet
routing not only randomises the defender's path but incorporates the
notion of {\em latency awareness} by choosing the route that minimises
the cumulative weight of (randomised) edge scores in the route. This
increases the likelihood of the defender meeting delivery deadlines as
compared to Markov chain routing. Since the random scores are only
known to the defender, and each defender uses different edge scores,
the attacker is unable to effectively predict the edge a defender may
be traveling.  As such Mixnet routing achieves a better balance
between the maximal adversarial uncertainty of Markov chain approach
vs the highly predictable shortest-path routing.

Variants of mixnet routing are possible that achieve a different
tradeoff between adversary uncertainty and latency. For instance, a
defender can follow (one of) the shortest path for most part but toss
a coin at intermediate nodes and either continue to follow the
shortest-path to the next hop, or undertake an $O(log n)$ random walk
and regenerate a shortest-path route from the end of the Markov chain
to the next destination on the tour. We will explore these variants in
future work.


\section{Evaluation}
\label{sec:results}
We consider 12 different road networks corresponding to popular cities
across the world where an automated real-time courier delivery system
might be financially viable. The security game described in
Section~\ref{sec:game} is played in a number of rounds. Each round
consists of an attack described in
Sections~\ref{sec:attacks}--~\ref{sec:attackslast}, when deployed
against each of the defense strategies in Section~\ref{sec:defenses}.


We simulate the interdiction game using real-world courier
workflows. Our analysis proceeds as follows. We initially investigate
the impact of each attack on the percentage of late deliveries when
shortest-path routing or a defense routing strategy is employed in a
multi-round game whilst also computing Nash equilibria. Nash
equilibrium is the solution to our adversarial game, in which an
attacker and defender choose a strategy while considering the
opponents choice, and neither benefits by changing their strategy.

Initially, we focus on the cities of London and Beijing for which we
have access to real courier traces. Subsequently, we validate our
findings at scale using graph data from ten other cities.

\paragraph*{Assumptions} We assume that the attacker has perfect information
about the road network including traffic information as this
information is publicly available. The defender also has access to
this information, however the defender is not aware of the attacker's
success rate on a particular route. This simulates the scenario that
defenders belong to different administrative domains (i.e no single
company owns them all) and hence their strategy selection is not
coordinated. Our goal is to understand the lower bound for adversary
success --- the best-case scenario for operators of fleets of
driverless vehicles. We assume that roads allow movement at the posted
speed limits. Consequently, our analysis is the most optimistic
scenario for developers of driverless vehicle technology, referred to
as the defender.

\subsection{London dataset}
\begin{figure}[!h]
    \centering
    \includegraphics[width=0.4\textwidth]{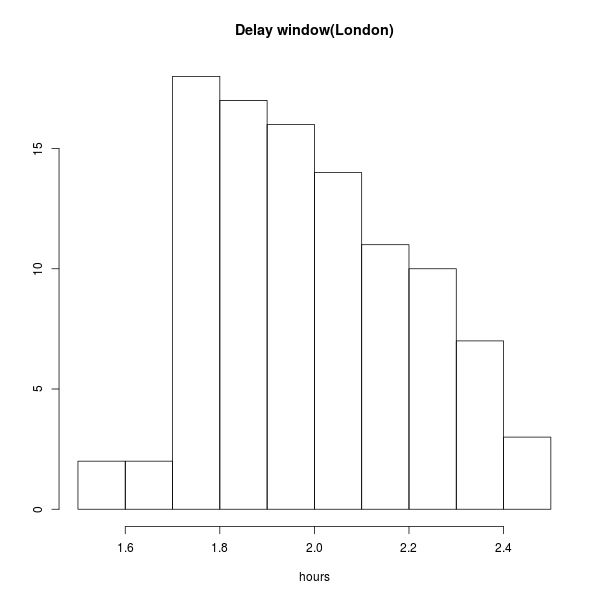}
    \caption{Delivery Window (London)}
    \label{fig:london:win}
\end{figure}

Our first dataset contains real courier traces for the city of London
provided by eCourier (\url{www.eCourier.co.uk}). eCourier provides
this data through the Open Street Map (OSM) project via a Creative
Commons license. This dataset contains traces of actual courier
movement over an eight week period in 2007 corresponding to half a
million deliveries. Each delivery is associated with a delivery window
which is a binary tuple composed of the earliest delivery time and the
latest acceptable delivery time for the item. We also obtained the
traffic and road maps for London via Open Street Map and generated a
road network graph. Figure~\ref{fig:london:win} shows the distribution
of the delivery windows and we can observe that the average delivery
window is about 2.2 hours. The average tour time in this dataset is
$\sim$11 hours. We assume that each successful ambush causes a delay of $M=10$ minutes. This is an optimistic estimate for the amount of time taken by a recovery vehicle to attend the scene of the attack and recover the vehicle to a new location, outside the attack zone. Longer delays would increase the delivery-failure rates experienced by couriers.

\begin{figure}[!h]
    \centering
    \includegraphics[width=0.5\textwidth]{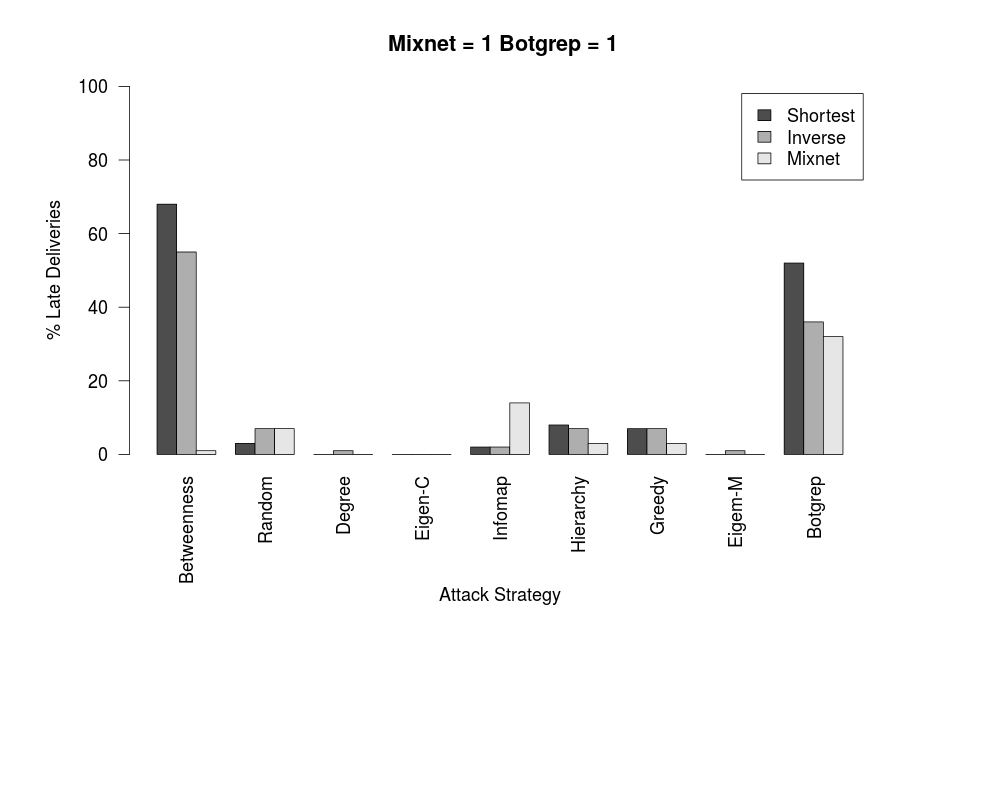}
    \caption{Late Deliveries vs. Attacks (London)}
    \label{fig:london:strategy}
\end{figure}

\paragraph*{Attacks vs. Defenses} We simulated the multi-round adversarial network interdiction game
over 83330 delivery schedules within this dataset.
Figure~\ref{fig:london:strategy}shows the impact on delivery time for
every combination of nine attack strategies and three defense
strategies, within a multi-round game involving thirty attackers (we
justify this in a future section).

Figure~\ref{fig:london:strategy} shows that most attacks on driverless
vehicles in London are effectively countered by shortest-path routing,
with the exception of the Betweenness and Botgrep attacks. Betweenness
attacks the edges which lie on the shortest-paths. Consequently, it
caused 70\% of deliveries to be late when the defender was using
shortest-path routing. When the defender switched to Mixnet routing,
there were no late deliveries, hence completely mitigating the
attack. Mitigation is achieved by the randomness of Mixnet routing,
which switches from using high-betweenness edges to leveraging edges
that are a part of mid to low conductance cuts, in order to route
efficiently. From Table~\ref{table:delay} we can see that 65\% of the
late deliveries caused by the Betweenness attack were critically
delayed (by $>$50\% of delay window, eg 1.1 hours for London) when
shortest-path routing is used. The Inverse defense strategy reduced
the amount of critical delays by 10\%, but Mixnet significantly
reduced the critical delays to 1\% of the overall late deliveries.
When the attacker switches strategies from Betweenness to Botgrep, the
Mixnet defense, unlike its effectiveness in defending against the
Betweenness attack, was less effective compared to using shortest-path
routing. The reason for this is that Botgrep (in common with other
modularity-based techniques) attacks low conductance cuts which are
crucial to Mixnet's routing efficiency, as they enable connectivity
between sparsely connected localities. Interestingly, the Inverse
centrality defense was the most successful at reducing the amount of
late deliveries caused by the Botgrep attack. Overall, the amount of
critical delays caused by Botgrep is less than Betweenness. Even
though Mixnet is less effective against Botgrep compared to
Betweenness, it still reduces the amount of critically late delays to
30\% of the overall late deliveries, which is lower than if
shortest-path routing or Inverse is used.

Next, we allowed attackers and defenders to adapt to eachother's
strategies. For the city of London, we found a pure Nash equilibrium
between {\em the Botgrep attack strategy and Mixnet defense
  strategy}. When adaptation is allowed, and the attacker employs the
Betweenness attack, the defender can deploy Mixnet to suitably counter
it. The attacker could counter the defender's move with the Botgrep
attack, maximising the attacker's payoff against Mixnet. However, no
other strategy increases the defender's payoff, and no other attack
improves the attacker's payoff, hence constituting a Nash equilibrium.

\begin{figure}[!h]
    \centering \includegraphics[width=.5\textwidth]{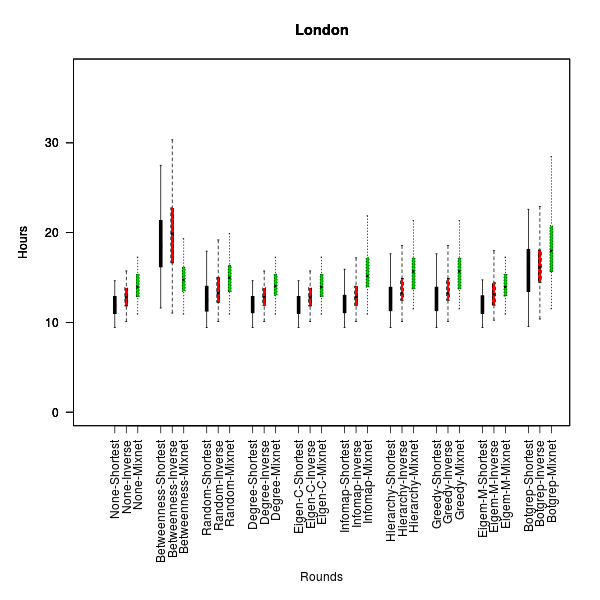}
    \caption{Tour Time vs. Attacks (London)}
    \label{fig:london:tour}
\end{figure}

\paragraph*{Impact of attacks on tour time} We also measured the increased
costs imposed by defences my measuring the numbers of hours a courier
would need to work for. With no attacks, the average working day is
$\sim$11 hours, shown by the length of an average tour when
shortest-path routing is used (Figure~\ref{fig:london:tour}). Even
where defenses are successful in minimising the impact of attacks by
ensuring the deliveries reach on time, the length of the workday
increases significantly which constitutes the extra cost of
resilience.

The Betweenness attack induces the largest increase in tour time
compared to all other attacks on this dataset. Although the Inverse
defense reduces the amount of late deliveries compared to
shortest-path routing shown in Figure~\ref{fig:london:strategy}, it
induces a higher tour length compared to shortest-path routing, with a
worst-case tour time of around 24 hours. Interestingly, the average
tour length when Mixnet is deployed against Betweenness is only
slightly more than when Mixnet routing is deployed under no attack; we
note however the the worst-case tour-length is significantly higher
under attack.

We also observed that Botgrep does not induce as many late deliveries
as the Betweenness attack, due to the overall tour time being
relatively similar. Although Mixnet reduces the amount of late
deliveries compared to Inverse and shortest-path routing for Botgrep,
as shown in Figure~\ref{fig:london:strategy}, it incurs a longer tour
time. Interestingly, the tour time incurred by Mixnet defense against
Botgrep is similar to the tour time for the Betweenness attack with
shortest-path routing, which caused the highest number of late
deliveries in this dataset.

\begin{figure}[!h]
    \centering
    \includegraphics[width=0.5\textwidth]{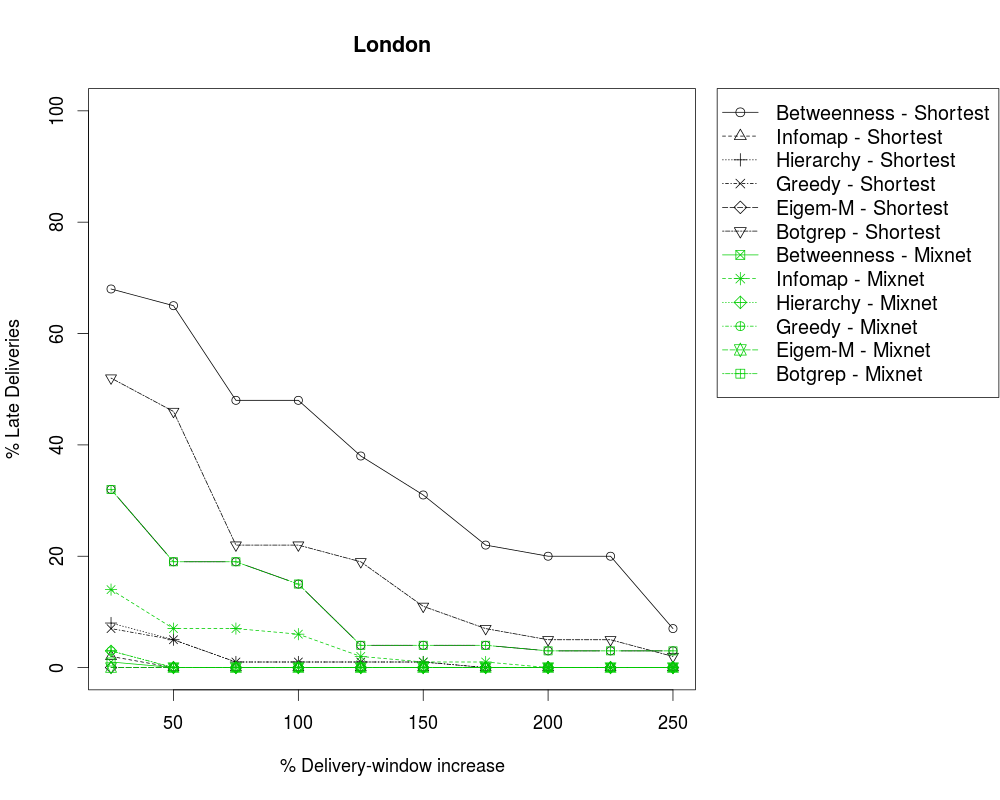}
    \caption{Late Deliveries vs. Delay Window Increase (London)}
    \label{fig:london:window}
\end{figure}

\paragraph*{Impact of delivery-window size} Next, we investigated whether
increasing the delivery window --- the buffer times available to a
courier before a delivery is classed as late --- would reduce the
number of late deliveries. The average delivery window size within the
dataset is 2.2 hours (Figure~\ref{fig:london:win}). The results of
increasing the delivery-window size are show in
Figure~\ref{fig:london:window}. Reductions in late deliveries start at
around a 75\% increase in the delivery window, which is 3.85
hours. Reducing the percentage of late deliveries to a serviceable
level of 5\% of total deliveries, requires significant increase in the
delay-window size, which has implications for the numbers of hours a
courier needs to work for to complete the day's work (or an increase
in the number of couriers). For example, in order to reduce late
deliveries from 75\% to 10\% for the Betweenness attack, this would
require a delivery window increase of around 250\% or around 5.5 hours
per delivery.

\begin{figure}[!h]
    \centering \includegraphics[width=0.5\textwidth]{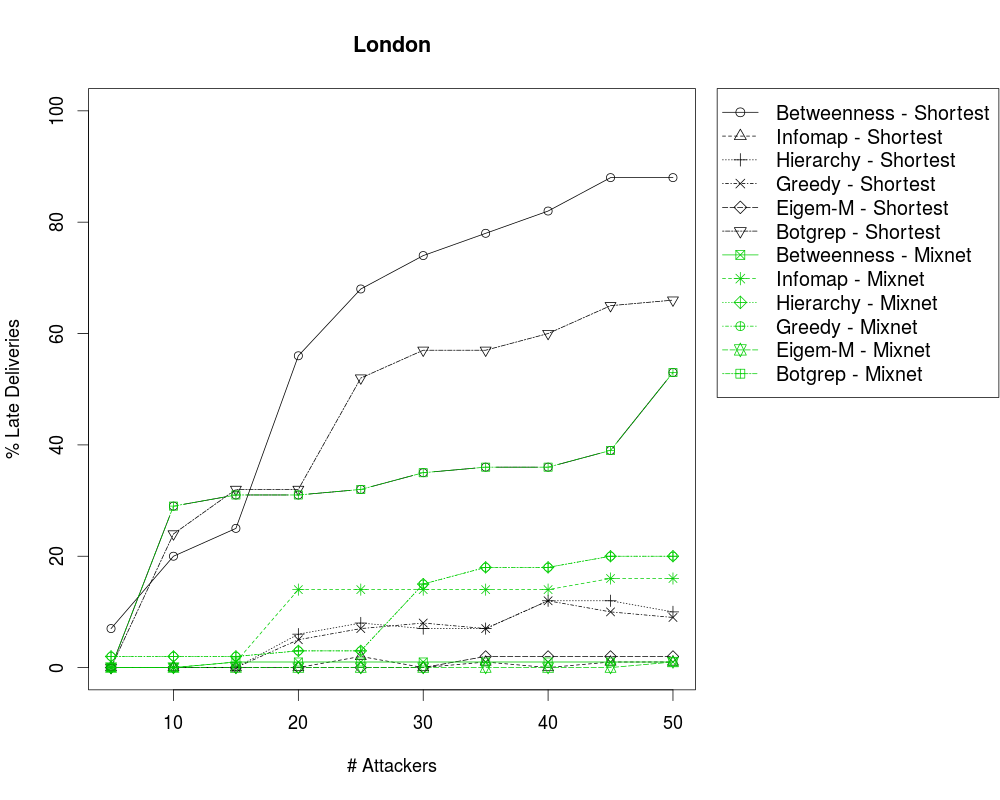}
    \caption{Late Deliveries vs. \# Attackers (London)}
    \label{fig:london:attackers}
\end{figure}

\paragraph*{Impact of attacker strength} To investigate the impact of
the number of attackers units on late deliveries, we controlled for
the number of attack resources available to the attacker. We assume
these attack units are coordinated by a single attacker who
coordinates the placement and strategies of all the attack units using
a command-and-control network. As shown in
Figure~\ref{fig:london:attackers}, as the number of attackers
increases, increasing numbers of edges get attacked which result in
increasing delivery times. We identified that on average, significant
increases in late deliveries occur between 10 and 30 attackers. As
well as this, attacks other than Betweenness and Botgrep show minimal
or no increases in late deliveries, regardless of how many attackers
are deployed. From these observations, we decided to run our previous
experiments with a baseline of 30 attackers.

\begin{table*}
\centering
\begin{tabular}{|ll|*{10}{c|}}
  \cline{3-12}
  \mc{1}{l}{}&&\mc{1}{|c|}{Lon}&\mc{1}{c|}{Bei}&\mc{1}{c|}{Bris}&\mc{1}{c|}{Bham}&\mc{1}{c|}{Cam}&\mc{1}{c|}{Gla}&\mc{1}{c|}{Eburgh}&\mc{1}{c|}{Del}&\mc{1}{c|}{Chi}&\mc{1}{c|}{Bos}\\
  \cline{3-12}
  \hline
    & Shortest & 65 & 94 & 85 & 96 & 100 & 94 & 81 & 98 & 34 & 97 \\
  \cline{3-12}
  Betweenness& Inverse & 55 & 99 & 88 & 98 & 100 & 88 & 82 & 97 & 40 & 72 \\
  \cline{3-12}
    & Mixnet & 1 & 10 & 25 & 2 & 75 & 6 & 1 & 3 & 1 & 0 \\
  \hline
    & Shortest & 3 & 22 & 18 & 9 & 54 & 15 & 7 & 7 & 5 & 13 \\
  \cline{3-12}
  Random& Inverse & 7 & 27 & 18 & 3 & 45 & 9 & 13 & 13 & 8 & 13 \\
  \cline{3-12}
    & Mixnet & 7 & 24 & 16 & 6 & 51 & 15 & 13 & 12 & 8 & 9 \\
  \hline
    & Shortest & 0 & 4 & 25 & 0 & 15 & 31 & 11 & 2 & 0 & 0 \\
  \cline{3-12}
  Degree& Inverse & 1 & 2 & 21 & 3 & 8 & 35 & 4 & 4 & 0 & 0 \\
  \cline{3-12}
    & Mixnet & 0 & 1 & 54 & 0 & 3 & 13 & 4 & 1 & 0 & 0 \\
    \hline
      & Shortest & 0 & 3 & 32 & 0 & 17 & 2 & 1 & 0 & 0 & 0 \\
    \cline{3-12}
    Eigen-C& Inverse & 0 & 2 & 42 & 0 & 12 & 5 & 0 & 10 & 0 & 0 \\
    \cline{3-12}
      & Mixnet & 0 & 1 & 77 & 0 & 9 & 9 & 66 & 14 & 0 & 0 \\
      \hline
        & Shortest & 0 & 38 & 31 & 1 & 58 & 10 & 20 & 12 & 4 & 45 \\
      \cline{3-12}
      Infomap& Inverse & 2 & 59 & 34 & 0 & 46 & 12 & 9 & 9 & 2 & 42 \\
      \cline{3-12}
        & Mixnet & 14 & 4 & 30 & 7 & 50 & 52 & 11 & 40 & 2 & 6 \\
        \hline
          & Shortest & 5 & 86 & 76 & 11 & 95 & 35 & 26 & 85 & 13 & 4 \\
        \cline{3-12}
        Hierarchy& Inverse & 7 & 77 & 69 & 23 & 94 & 31 & 23 & 83 & 14 & 5 \\
        \cline{3-12}
          & Mixnet & 3 & 54 & 62 & 4 & 83 & 50 & 27 & 68 & 38 & 3 \\
          \hline
            & Shortest & 5 & 86 & 76 & 11 & 95 & 35 & 26 & 85 & 13 & 4 \\
          \cline{3-12}
          Greedy& Inverse & 7 & 77 & 69 & 23 & 94 & 31 & 23 & 83 & 14 & 5 \\
          \cline{3-12}
            & Mixnet & 3 & 54 & 62 & 4 & 83 & 50 & 27 & 68 & 38 & 4 \\
            \hline
              & Shortest & 0 & 72 & 2 & 3 & 82 & 0 & 32 & 39 & 2 & 0 \\
            \cline{3-12}
            Eigen-M& Inverse & 1 & 74 & 3 & 5 & 81 & 2 & 27 & 30 & 5 & 2 \\
            \cline{3-12}
              & Mixnet & 0 & 19 & 2 & 2 & 80 & 1 & 15 & 60 & 5 & 6 \\
              \hline
                & Shortest & 46 & 68 & 76 & 14 & 83 & 19 & 59 & 64 & 26 & 3 \\
              \cline{3-12}
              Botgrep& Inverse & 36 & 74 & 77 & 15 & 82 & 27 & 48 & 57 & 18 & 5 \\
              \cline{3-12}
                & Mixnet & 30 & 21 & 73 & 13 & 89 & 47 & 44 & 9 & 62 & 1 \\
              \hline
\end{tabular}
\caption{\% of Critically Delayed Late Deliveries}
\label{table:delay}
\end{table*}

\subsection{Beijing dataset} 
\begin{figure}[!h]
    \centering
    \includegraphics[width=0.4\textwidth]{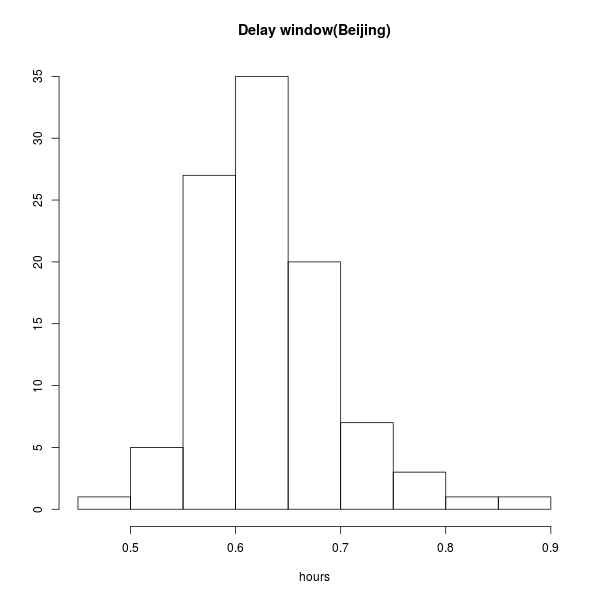}
    \caption{Delivery Window (Beijing)}
    \label{fig:beijing:win}
\end{figure}

Our second dataset for the city of Beijing is
two orders of magnitude larger than the London courier dataset. This
dataset contains traces generated by 30000 couriers over a period of
three months between making a combined total of $75$ million
deliveries. Figure~\ref{fig:beijing:win} shows the distribution of the
delivery windows and we can observe that the average delivery window
is about 30 minutes. The average tour time in this dataset is $\sim$4
hours, owing to a large number of couriers being engaged for a
fraction of a working day.

\begin{figure}[!h]
    \centering
    \includegraphics[width=0.5\textwidth]{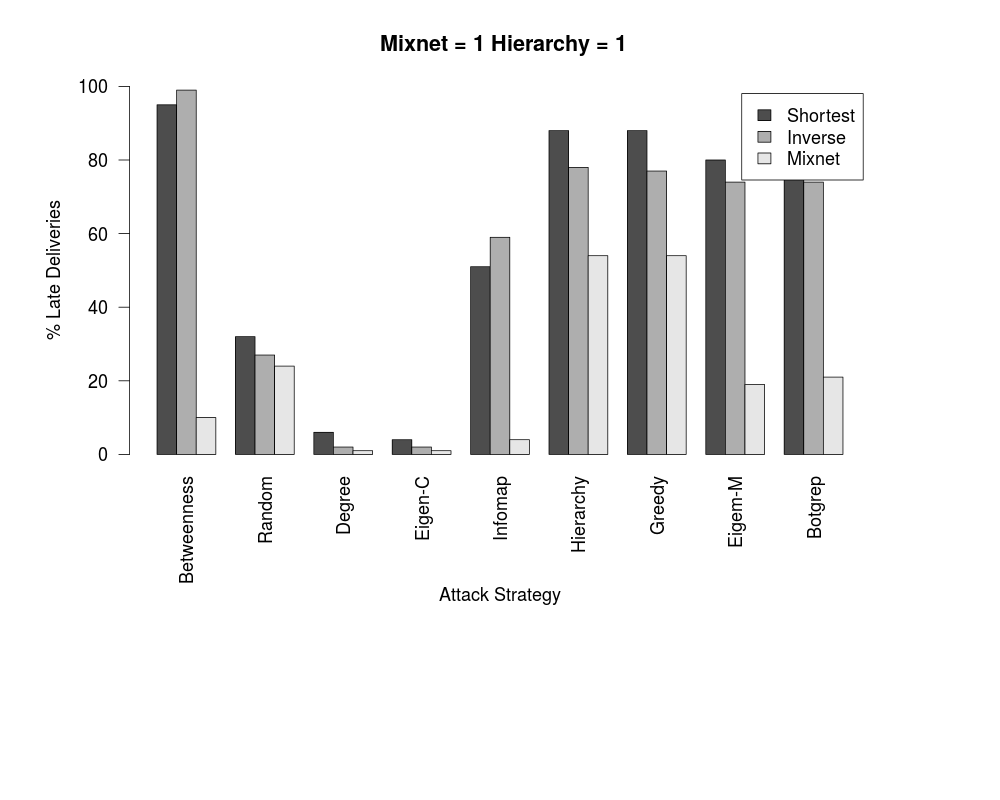}
    \caption{Late Deliveries vs. Attacks (Beijing)}
    \label{fig:beijing:strategy}
\end{figure}

\paragraph*{Attacks vs. Defenses} Figure~\ref{fig:beijing:strategy} shows that most attacks had some degree of impact on the amount of late
deliveries when shortest-path routing was used. Specifically, the
Betweenness, Eigen-modularity and Botgrep attacks were successful in
inducing high rates of late deliveries. The Betweenness attack is
effectively mitigated by the Mixnet defense strategy, similar to
London. Unlike London, the Inverse defense strategy incurs a higher
percentage of late deliveries when employed against the Betweenness
attack than if shortest-path routing is used. This is because London
has higher redundancy in terms of the number of disjoint shortest
paths within the road network. With an increased number of attackers
running the betweenness attack, London can be expected to show a
similar trend i.e Inverse performs worse than shortest-path routing.
This does not mean there is not enough redundancy between
origin-destination pairs in Beijing, however routing techniques based
on shortest-paths cannot locate such routes. Mixnet routing however
can do so. Similar to the London dataset, the Mixnet strategy also
significantly reduces the amount of critical delays. This indicates
the importance of leveraging low-conductance paths rather than
shortest-path routing to construct better defences.

In both London and Beijing, attacks leveraging betweenness centrality
and low-conductance cuts are fairly successful, while Mixnet is the
only serviceable defense. Table~\ref{table:delay}, shows that the
Betweenness attack causes 94\% of late deliveries to be critically
delayed and 99\% of these delays to be critical when the Inverse
defense is employed. Modularity-based attacks also induce high
percentages of critically delayed late deliveries. High percentages of
critical delays are induced by the majority of attacks with the
exception of Degree and Eigen-Centrality, which is unlike the London
dataset where high percentages of critically delayed late deliveries
are only induced by the Betweenness and Botgrep attacks. Modularity
attacks target low conductance cuts which Mixnet uses to improve
routing efficiency. We expect to observe a higher amount of late
deliveries when Mixnet is deployed against modularity-based
attacks. This is demonstrated by the Hierarchy and Greedy attacks in
Figure~\ref{fig:beijing:strategy}, with Mixnet incurring a lower
percentage of late deliveries compared to the Inverse defense. In
regards to conductance-based attacks, we find that Botgrep is a
successful attack in London whereas Hierarchical-modularity is the
most successful attack in Beijing. 

As with London, we also found a pure Nash equilibrium in Beijing between the
{\em Hierarchical-modularity} attack and {\em Mixnet} defense.


%

%
 %

\begin{figure}[!h]
 \centering
 \includegraphics[width=0.5\textwidth]{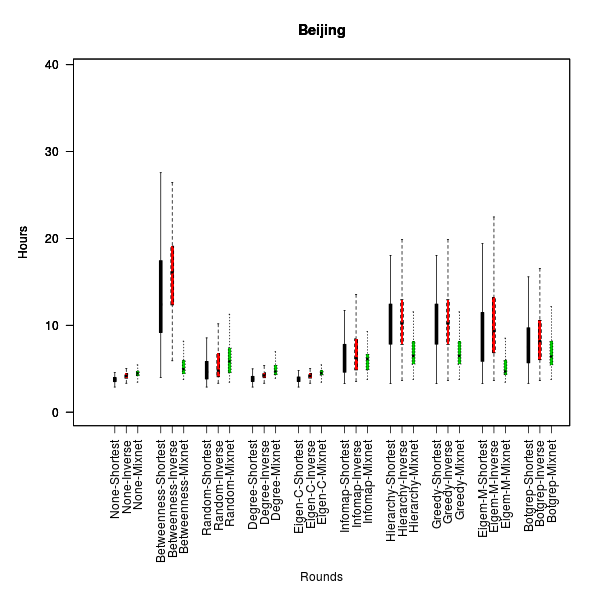}
 \caption{Tour Time vs. Attacks (Beijing)}
 \label{fig:beijing:tour}
\end{figure}

\paragraph*{Impact of attacks on tour time} Next, we investigated the impact of the different attack strategies on the tour time of couriers. From Figure~\ref{fig:beijing:tour} we can observe that the average tour time for a working day is $\sim$4 hours, shown by the length of an average tour when shortest-path routing is used with no attack. Only a minimal increase in tour time is observed when defense strategies are employed with no attack strategy used.
Similar to the London dataset, the Betweenness attack significantly increases the tour time of a courier to around 11 hours when no defensive-routing strategy is used. The Inverse defense further increases this, incurring a tour time of around 18 hours. Mixnet performs well to keep the tour time low, showing only a small increase compared to when it is employed against no attack strategy.
From Figure~\ref{fig:beijing:tour}, we can also identify that on average modularity-based attacks significantly increase the tour time compared to the average tour time by at least 50\%. Mixnet however manages to reduce the tour time incurred by the modularity-based attacks by around the same amount.

\begin{figure}[!h]
    \centering
    \includegraphics[width=0.5\textwidth]{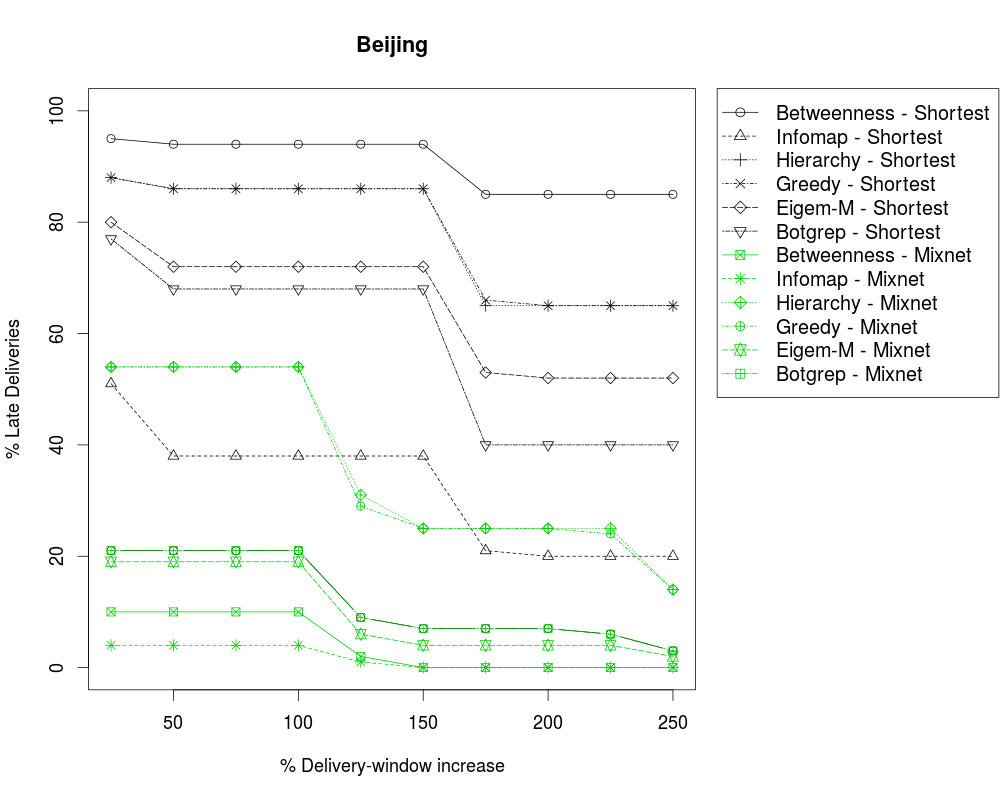}
    \caption{Late Deliveries vs. Delay Window Increase (Beijing)}
    \label{fig:beijing:window}
\end{figure}

\paragraph*{Impact of variable delivery windows} The next experiment we ran on this dataset was to investigate the impact of variable delivery windows on the amount of late deliveries. Figure~\ref{fig:beijing:window} shows us that significant decreases in the amount of late deliveries only occur after at least a 100\% in the existing delivery window. From Figure~\ref{fig:beijing:win}, we identified that the average delivery window was 30 minutes for this dataset. From our previous observation, we can deduce that a substantial reduction in late deliveries will be seen with a delivery window of about one hour. For an ideal amount of late deliveries, such as around 10\% like the London dataset, an increase of at least 250\% to the delay window (2.5 hours) is required. However, we can observe that attacks such as Betweenness still incur a very high percentage of late deliveries even with a 250\% increase in the delivery window. This suggests that increasing the delivery window alone does not resolve the attack.

\begin{figure}[!h]
    \centering
    \includegraphics[width=0.5\textwidth]{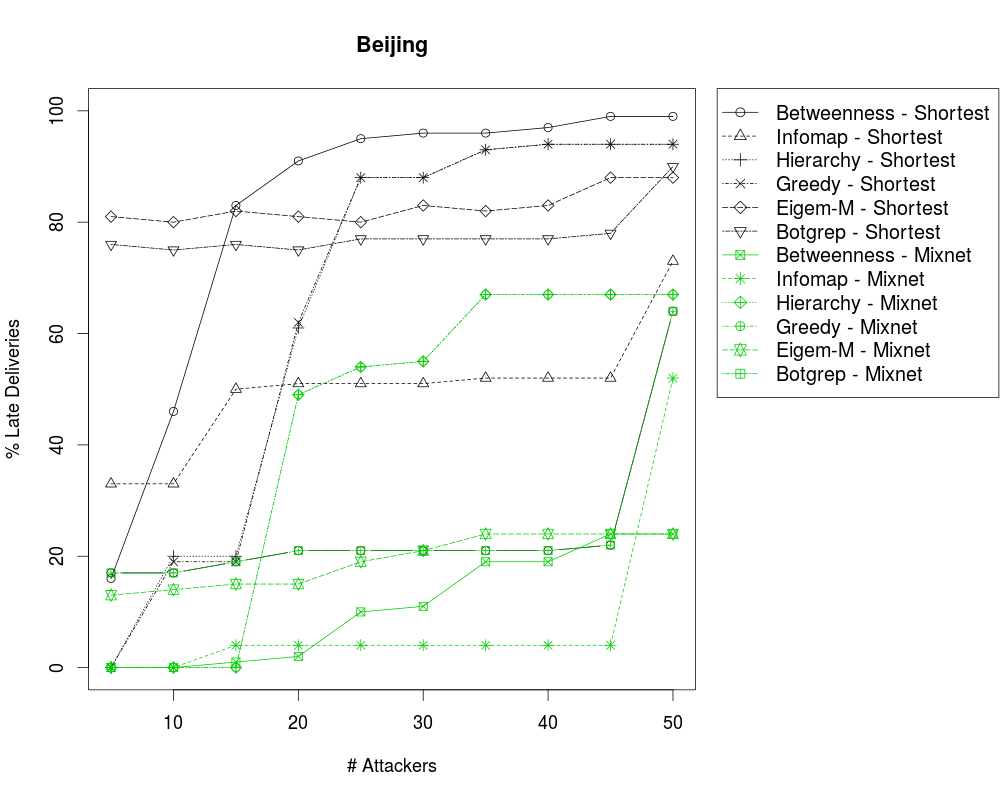}
    \caption{Late Deliveries vs. \# Attackers (Beijing)}
    \label{fig:beijing:attackers}
\end{figure}

\paragraph*{Impact of the number of attackers} Our final experiment on this dataset was to investigate the impact of the number of attackers on late deliveries. From Figure~\ref{fig:beijing:attackers} we can see that on average, there are significant increases between 1 and 30 attackers. The baseline is derived by observing how many attackers are required to induce significant failure rates using any attack with shortest-path routing. We observe that Botgrep, Eigen-Modularity, Greedy modularity, and Betweenness are very successful even at a fairly low attacker count of 5--10 attackers. However, to keep our experiments consistent with the London dataset for comparison, we decided to use the same baseline of 30 attackers, as it covers the significant increases in late deliveries for attacks on averages.

\subsection{Synthetic dataset} Our third dataset, is composed of synthetic
courier traces combined with real road network data generated via OSM
data for the following cities: Birmingham (UK),
Boston (USA), Bristol (UK), Cambridge (UK), Chicago
(USA), Delhi (India), Edinburgh (UK) and Glasgow (UK). The purpose of this dataset is to
expand our analysis beyond London and Beijing. To generate synthetic
traces, we use the London database as a basis. The number of couriers
are maintained but the locations are randomised in a
distance-preserving manner i.e the distance between consecutive
locations is identical on both the synthetic and real job cards for
any courier. In the London dataset, each courier has a job card
created on a per-day basis, that lists the delivery locations, times,
and a delivery window which serves as buffer to indicate the maximum
possible lateness allowed. We replace the first location on the card
with a location from the city of interest, chosen uniformly at
random. The subsequent locations on the card are replaced with another
random location such that the travel time between consecutive
locations on the synthetic card is the same as real job card. The
delivery window size (difference between the latest possible delivery
time and the window start time) is also maintained the same as the
London dataset. A job is marked late if it is delivered beyond the
maximum allowable delivery period.

\begin{figure*}[!ht]
 \vspace{-2em}
 \centering
   \vspace{-2em}
  \subfigure[Bristol, UK]{\label{fig:synthetic:strategy:bri}{\includegraphics[width=3.0in]{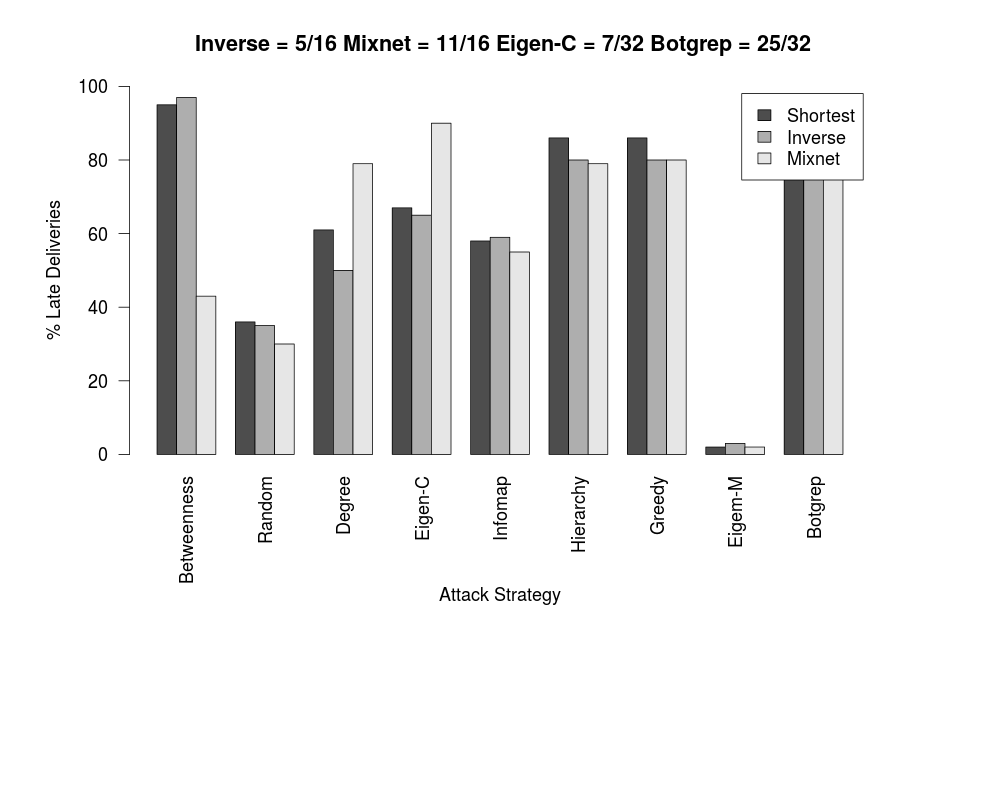}}}
  \subfigure[Birmingham, UK]{\label{fig:synthetic:strategy:bir}{\includegraphics[width=3.0in]{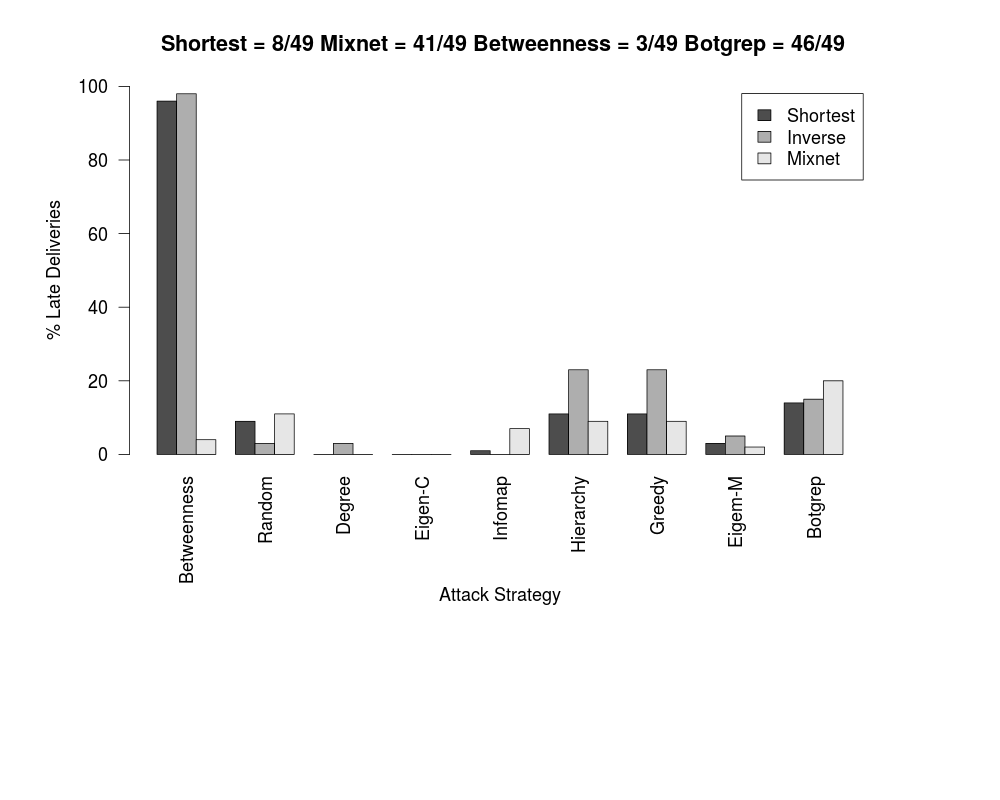}}}
  \vspace{-2em}
  \subfigure[Cambridge, UK]{\label{fig:synthetic:strategy:cam}{\includegraphics[width=3.0in]{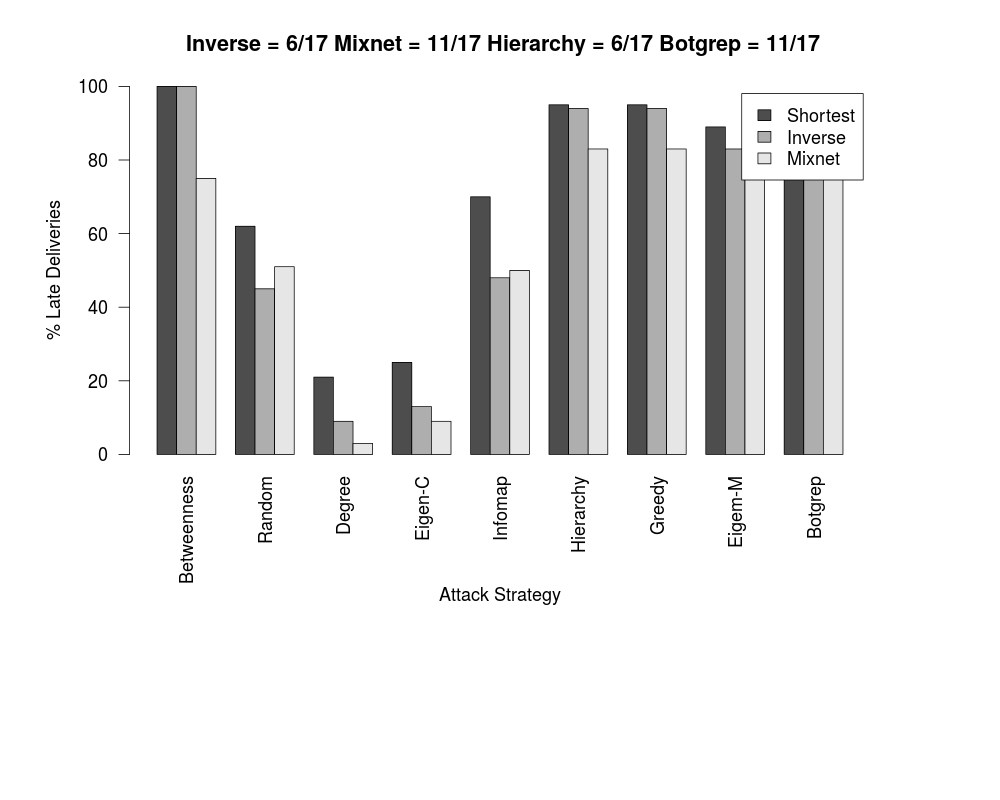}}}
  \subfigure[Glasgow, UK]{\label{fig:synthetic:strategy:gla}{\includegraphics[width=3.0in]{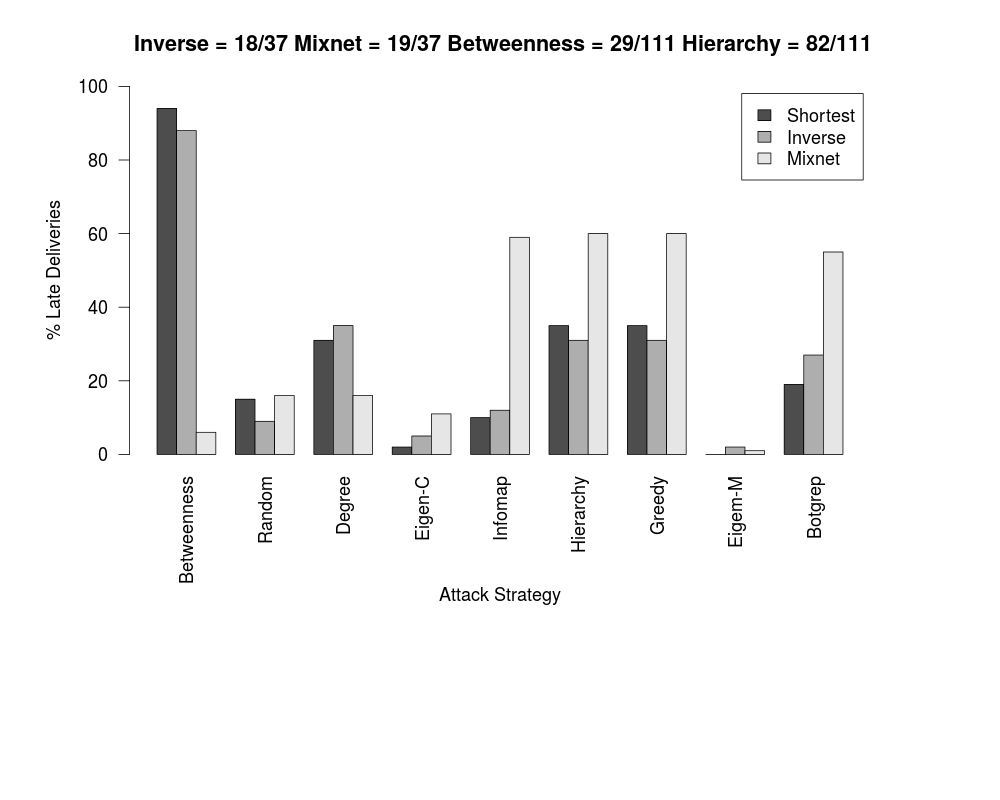}}}
  \vspace{-2em}
  \subfigure[Edinburgh, UK]{\label{fig:synthetic:strategy:edi}{\includegraphics[width=3.0in]{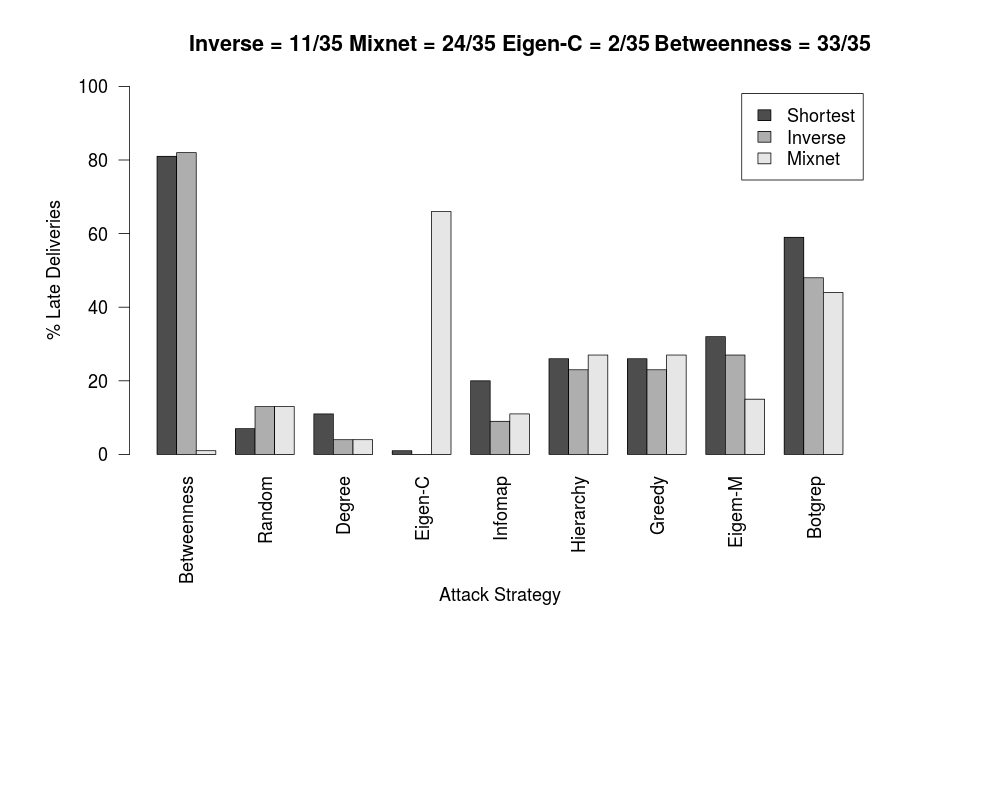}}}
  \subfigure[Delhi, India]{\label{fig:synthetic:strategy:del}{\includegraphics[width=3.0in]{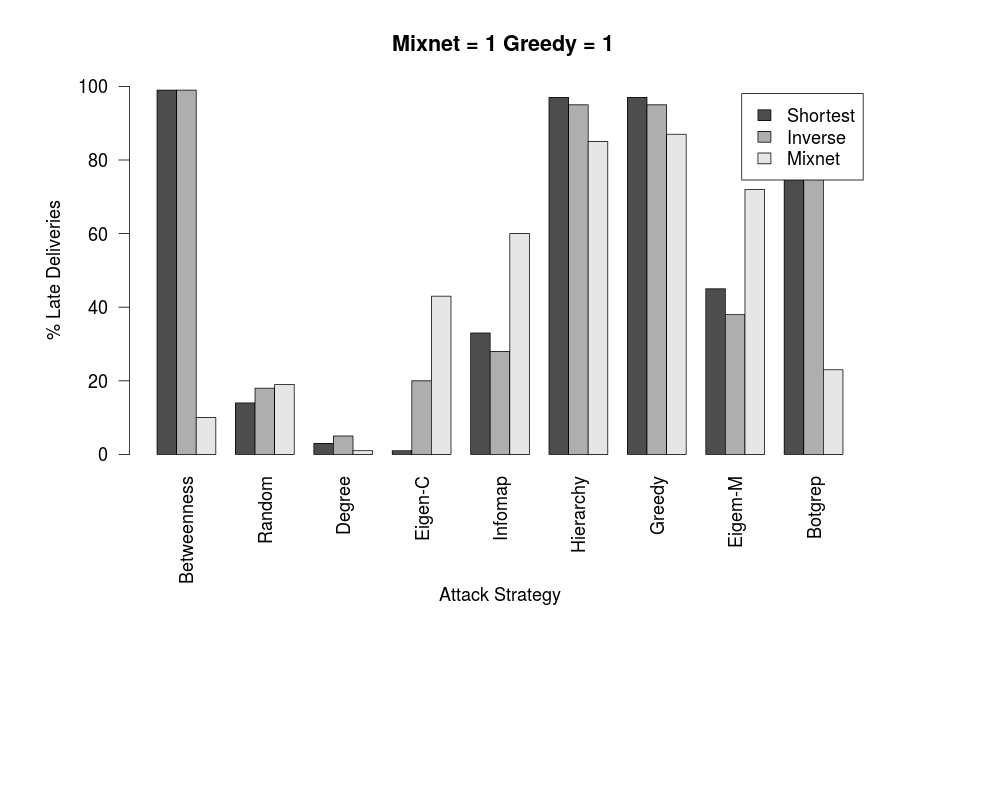}}}
    \vspace{-2em}
  \subfigure[Chicago, USA]{\label{fig:synthetic:strategy:chi}{\includegraphics[width=3.0in]{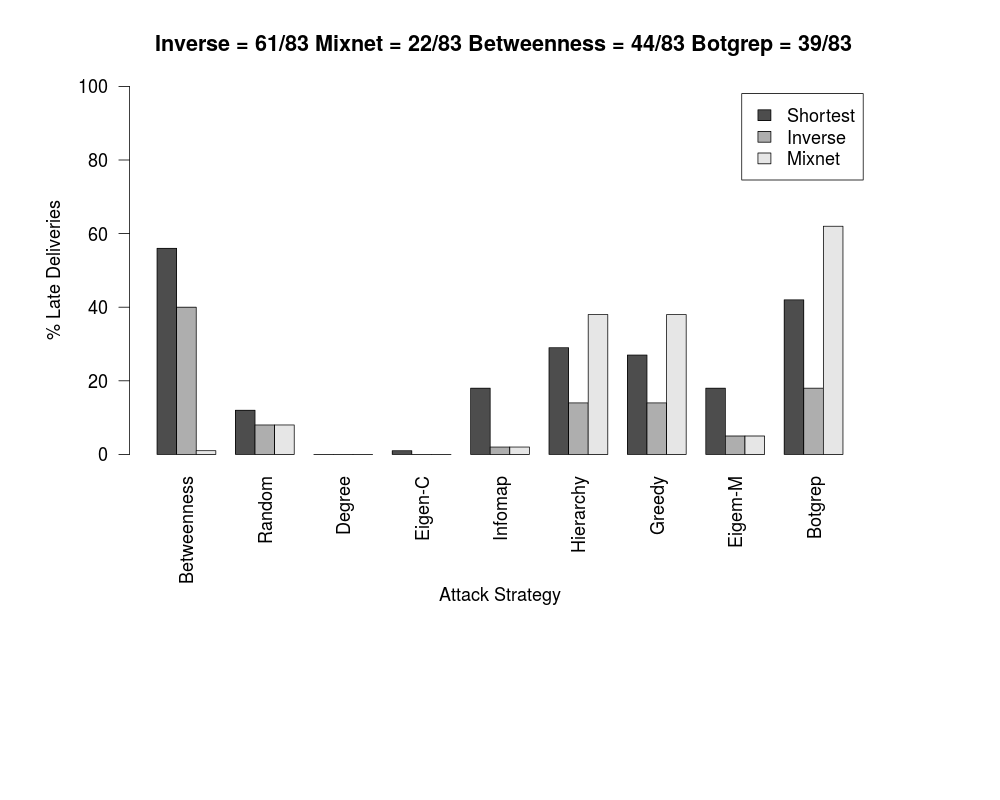}}}
    \subfigure[Boston, USA]{\label{fig:synthetic:strategy:bos}{\includegraphics[width=3.0in]{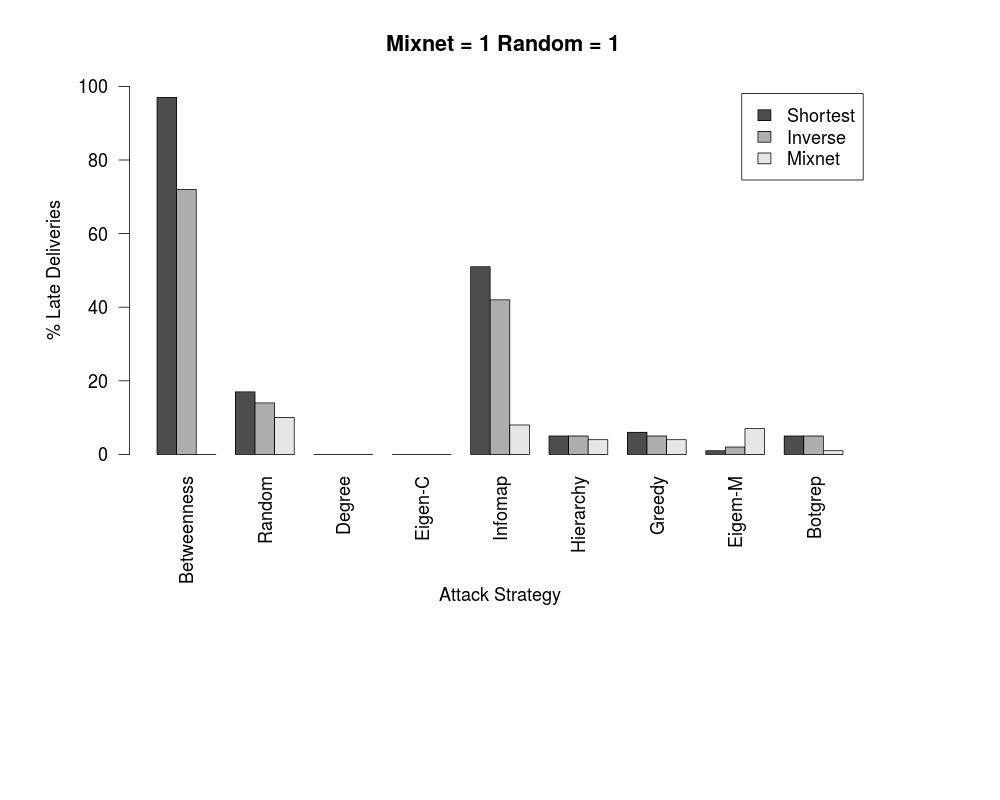}}} 
  \caption{Effect on \% of Late Deliveries for Attack Strategies}
  \label{fig:synthetic:strategy}
\end{figure*}

\paragraph*{Impact of Attack and Defense Strategies} We first evaluated the effectiveness of the different attack strategies on the percentage of late deliveries for each of the cities in our synthetic dataset, as well as the effectiveness of these attacks when a defense strategy is employed, with the results shown in Figure~\ref{fig:synthetic:strategy}. Overall, we observed that the Betweenness attack was the most successful attack on all the synthetic traces. For the Bristol, Birmingham and Edinburgh datasets, we noticed that the Inverse defense incurred a higher percentage of late deliveries compared to shortest-path routing, the same outcome shown in our results for Beijing.
In our previous datasets, we observed that modularity-based attacks caused a large number of datasets but were mitigated by Mixnet. We noticed that for some cities in our synthetic dataset, such as Glasgow (Figure~\ref{fig:synthetic:strategy:gla}), Mixnet performs exceedingly worse  as low-conductance edges across dense clusters are identified and targeted by modularity-based attacks.

%
%
%
%

Similar to London, our results on the Birmingham dataset in Figure~\ref{fig:synthetic:strategy:bir} show that the majority of attacks are not that effective. For example, the degree centrality attack has no impact on late deliveries even when no defense is employed.
Interestingly, for most cities in our synthetic dataset, Eigen-Centrality attacks have little or no impact on late deliveries. More specifically, our results for Edinburgh in Figure~\ref{fig:synthetic:strategy:edi} show that Eigen-Centrality with shortest-path routing incurs a very small number of late deliveries, but Mixnet causes $\sim$70\% of late deliveries. To investigate this further, we looked at the number of critically delayed late deliveries for these cities shown in Table~\ref{table:delay}. We identified that Mixnet causes 66\% of the late deliveries for the Eigen-Centrality attack on Edinburgh to be critically delayed. Overall, we observed that for all cities in our synthetic dataset, with the exception of Chicago and Boston, the modularity-based attacks incur the highest amounts of critical delays. The results from the table do show that for all cities, Mixnet reduces the amount of critical delays --- however, not substantially. This means that while Mixnet is able to mitigate the attack to some extent, these cities have relatively lower numbers of low-conductance cuts ($\rho < 0.076$) across localities which restricts the number of redundant paths available to Mixnet whilst under attack.

\begin{figure*}[!ht]
 \vspace{-2em}
 \centering
   \vspace{-2em}
  \subfigure{\label{fig:synthetic:tour:bri}{\includegraphics[width=2.25in]{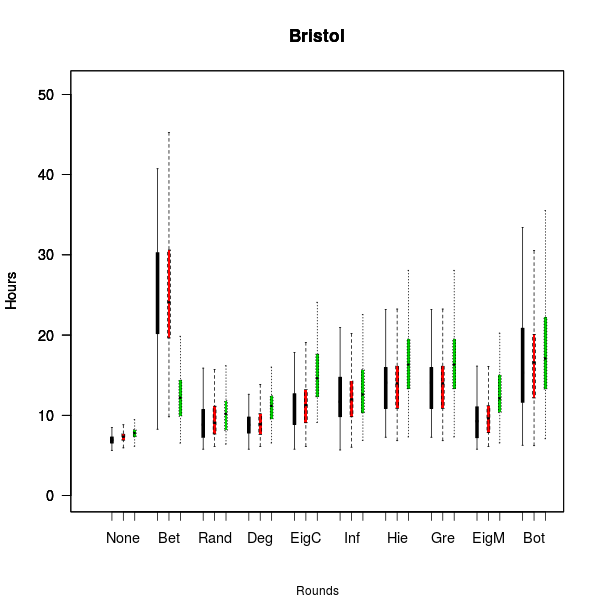}}}
  \subfigure{\label{fig:synthetic:tour:bir}{\includegraphics[width=2.25in]{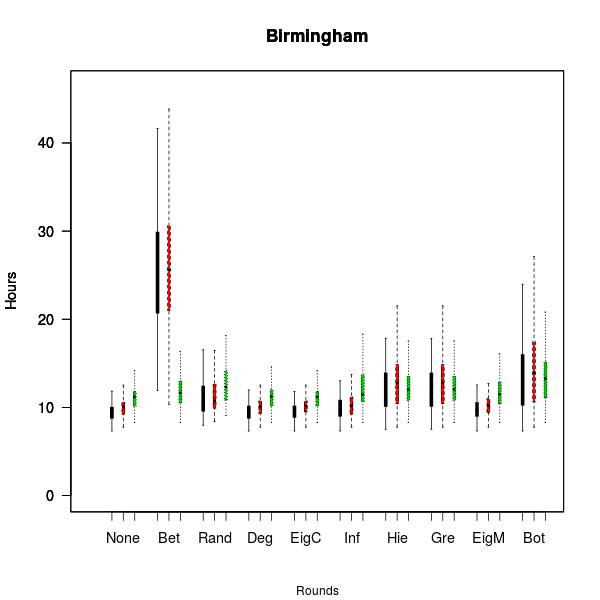}}}
  \subfigure{\label{fig:synthetic:tour:cam}{\includegraphics[width=2.25in]{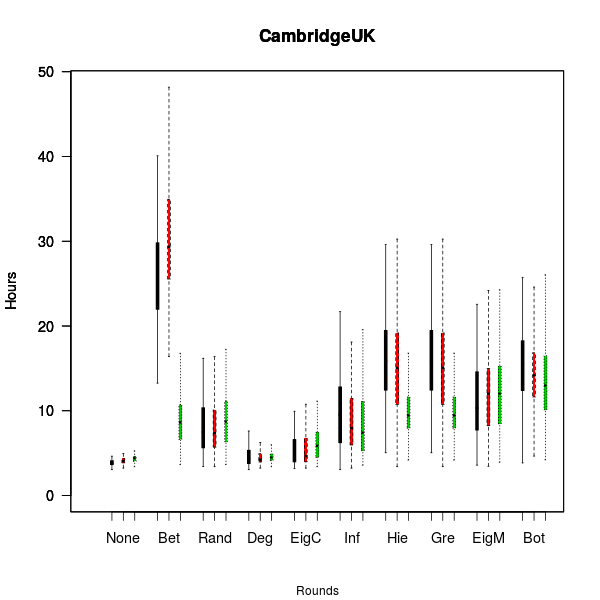}}}
  \subfigure{\label{fig:synthetic:tour:gla}{\includegraphics[width=2.25in]{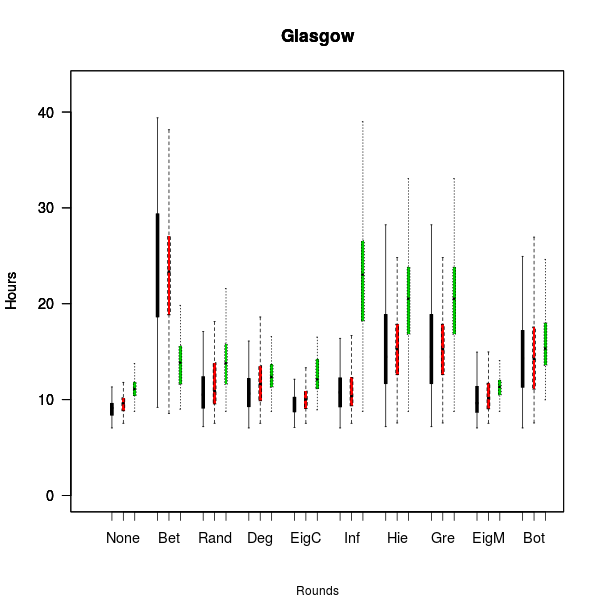}}}
  \subfigure{\label{fig:synthetic:tour:edi}{\includegraphics[width=2.25in]{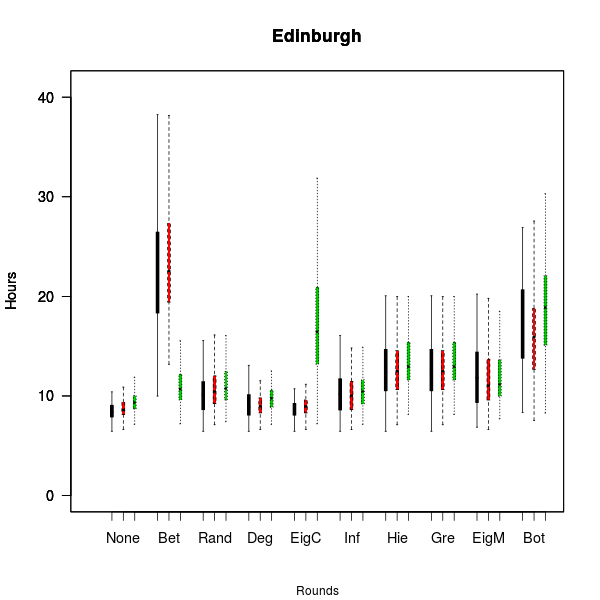}}}
  \subfigure{\label{fig:synthetic:tour:del}{\includegraphics[width=2.25in]{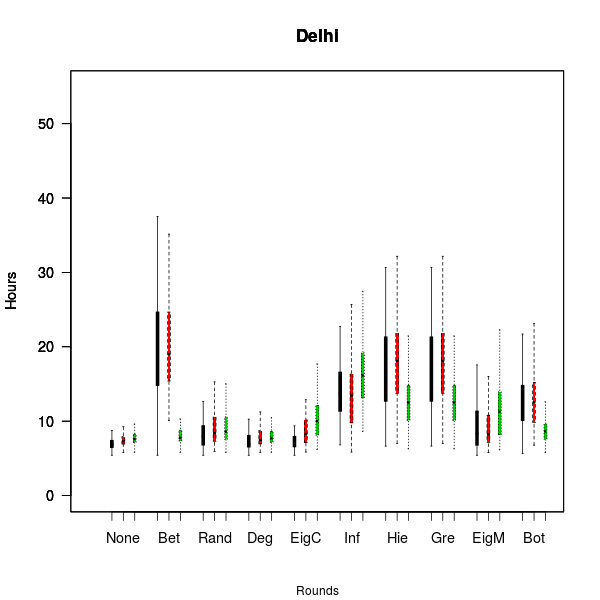}}}
  \subfigure{\label{fig:synthetic:tour:chi}{\includegraphics[width=2.25in]{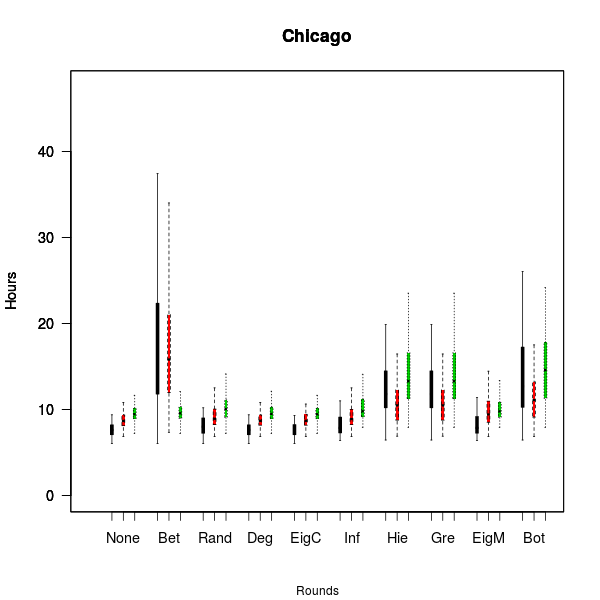}}}
    \subfigure{\label{fig:synthetic:tour:bos}{\includegraphics[width=2.25in]{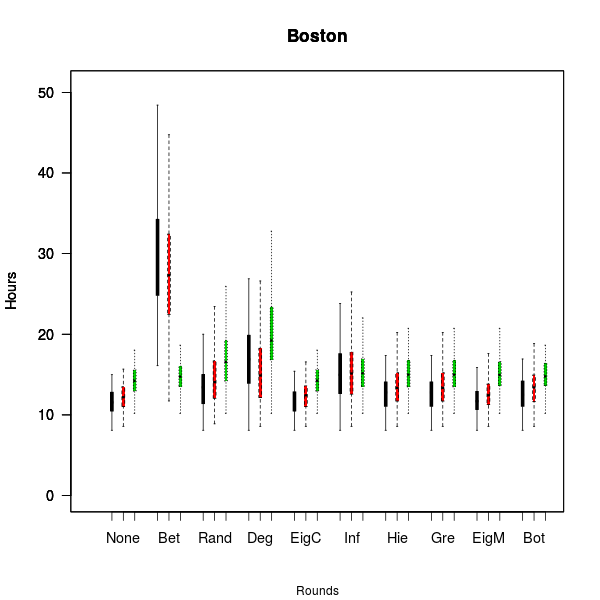}}}
    \subfigure{\label{fig:synthetic:tour:leg}{\includegraphics[width=2.25in]{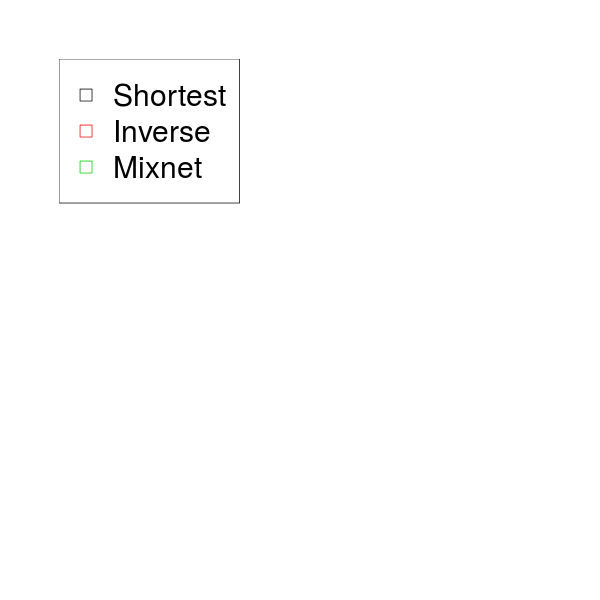}}}

    \caption{Tour Time vs. Attacks}
  \label{fig:synthetic:tour}
\end{figure*}

\paragraph*{Effect of Attack Strategies on Tour Time} Our next experiment on the synthetic dataset was to investigate the impact on attack strategies on tour time. Figure~\ref{fig:synthetic:tour} shows the effect of attacks on tour time. Overall, we observed that for most cities, the average tour time for a working day is between 8 and 10 hours. Boston has a slightly higher average tour time of $\sim$12 hours. The average tour time is shown by the length of an average tour when shortest-path routing is used with no attack.
Interestingly, we identified that increases in tour time correlates with the percentage of late deliveries shown in Figure~\ref{fig:synthetic:strategy}. For example, in Figure~\ref{fig:synthetic:strategy:edi} our results show that Mixnet incurs a high amount of late deliveries when deployed against Eigen-Centrality. In Figure~\ref{fig:synthetic:tour:edi} we can see the same increase in tour time when Mixnet is used, with the tour time increasing from $\sim$10 hours to nearly 20 hours.
The Betweenness attack also incurs the highest tour time for all cities, with Mixnet effectively reducing the tour time as well as the amount of late deliveries.

\begin{figure*}[!ht]
 \vspace{-2em}
 \centering
   \vspace{-2em}
  \subfigure{\label{fig:synthetic:window:bri}{\includegraphics[width=2.0in]{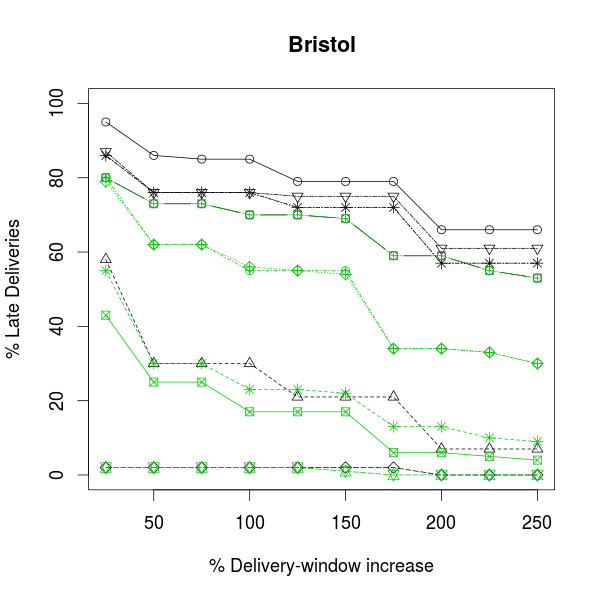}}}
  \subfigure{\label{fig:synthetic:window:bir}{\includegraphics[width=2.0in]{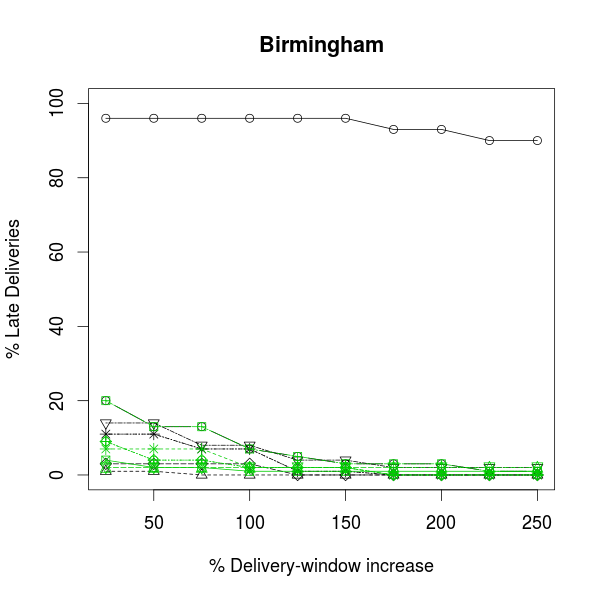}}}
  \subfigure{\label{fig:synthetic:window:cam}{\includegraphics[width=2.0in]{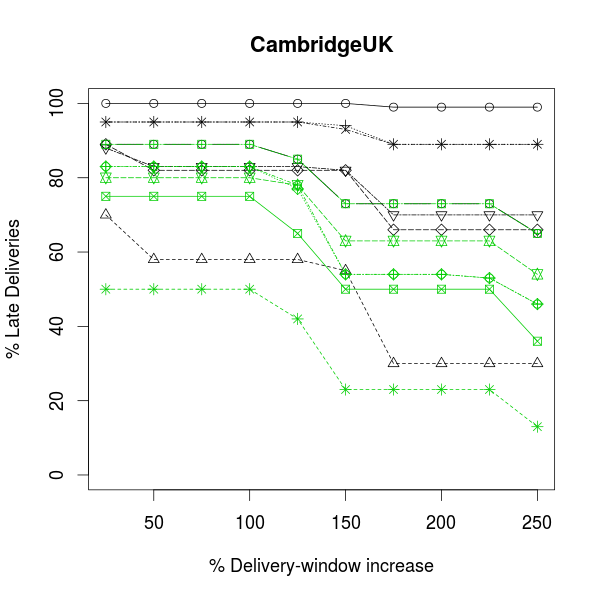}}}
  \subfigure{\label{fig:synthetic:window:gla}{\includegraphics[width=2.0in]{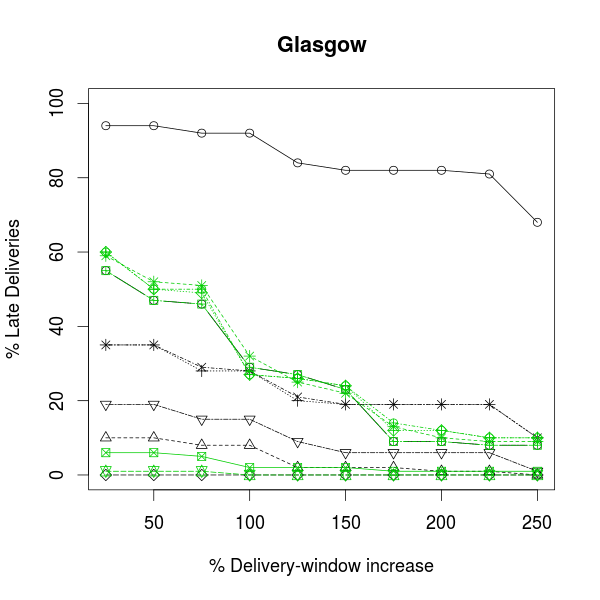}}}
  \subfigure{\label{fig:synthetic:window:edi}{\includegraphics[width=2.0in]{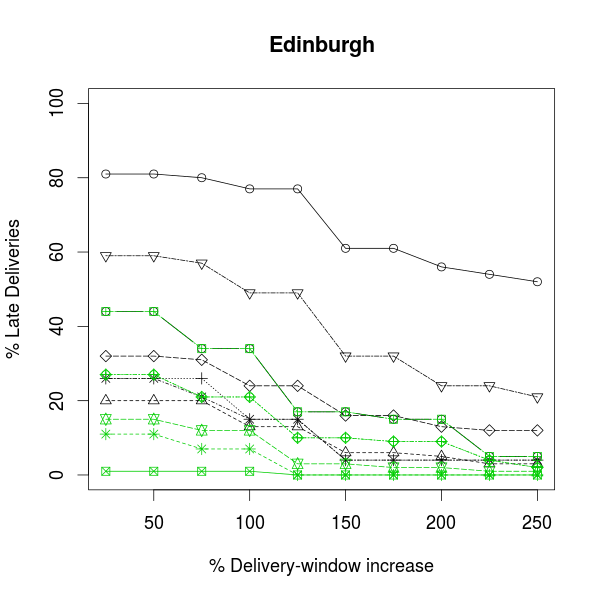}}}
  \subfigure{\label{fig:synthetic:window:del}{\includegraphics[width=2.0in]{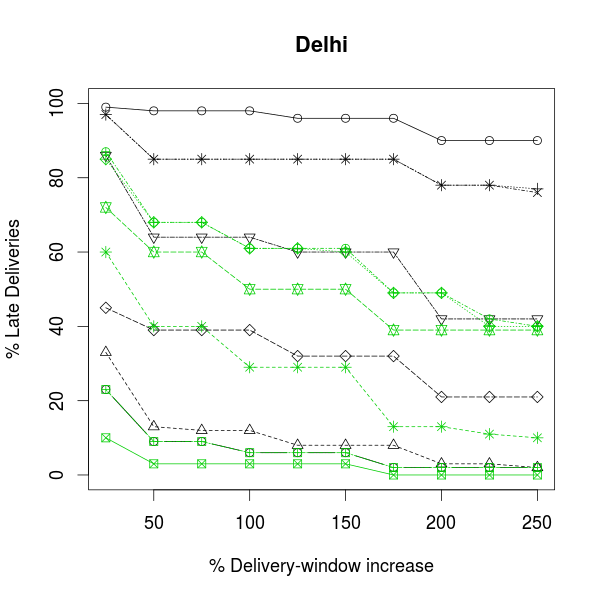}}}
  \subfigure{\label{fig:synthetic:window:chi}{\includegraphics[width=2.0in]{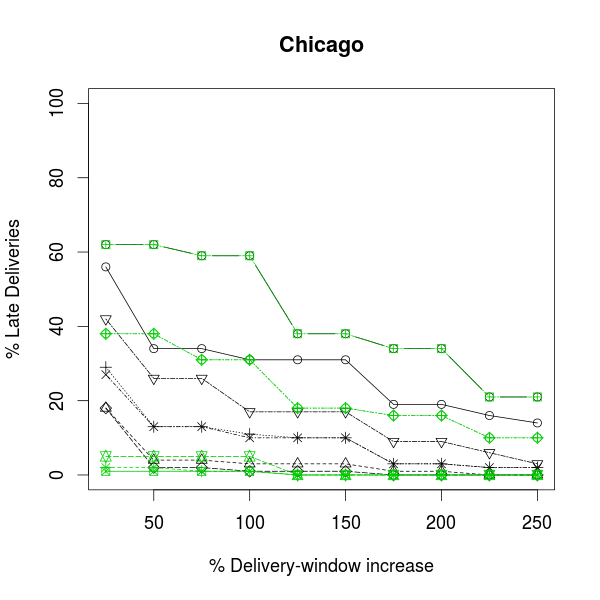}}}
  \subfigure{\label{fig:synthetic:window:bos}{\includegraphics[width=2.0in]{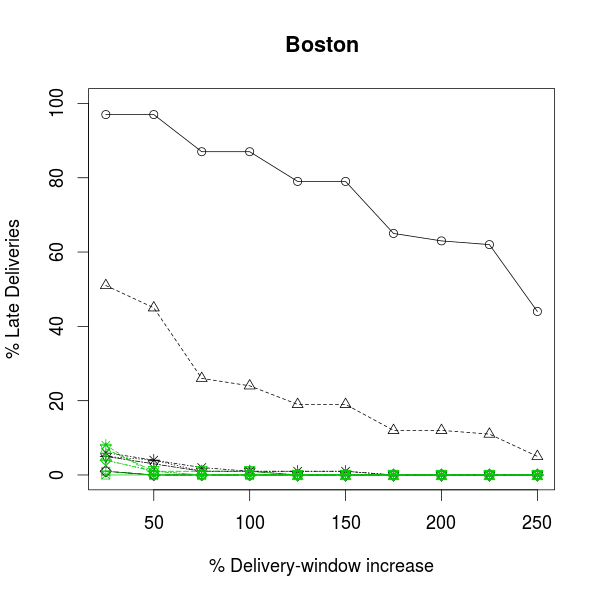}}}
  \subfigure{\label{fig:synthetic:window:leg}{\includegraphics[width=2.0in]{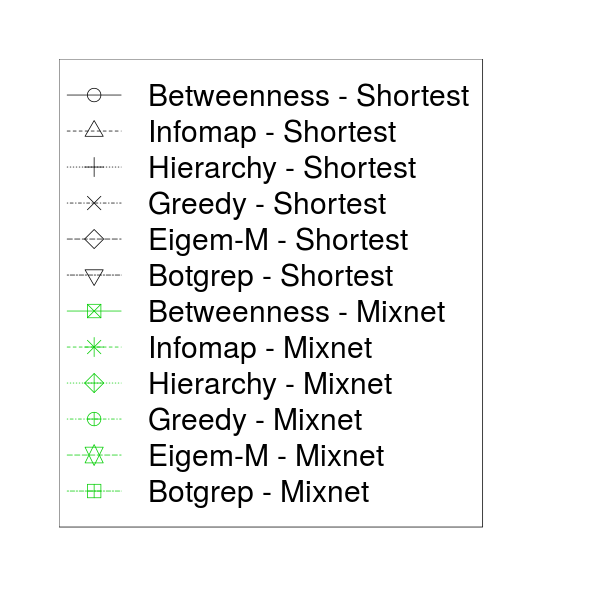}}}

  \caption{Late deliveries vs Delivery Window Increase}
  \label{fig:synthetic:window}
\end{figure*}

\paragraph*{Impact of Delivery Window Size} Our final experiment on this
dataset was to investigate the impact of the size of delivery windows
on the amount of late deliveries. As previously described, the
synthetic dataset is based off the London dataset such that the
delivery window size is maintained the same. Therefore we can state
that for all cities in our synthetic dataset, the average delivery
window is 2.2 hours. The results of this experiment are shown in
Figure~\ref{fig:synthetic:window}.  For all cities we observed that
the betweenness attack, regardless of the defense strategy, incurred
the highest percentage of late deliveries even with an increase in the
delivery window. For all cities, we noticed that substantial decreases
in late deliveries only occur after in increase in delivery window of
around 100\% (4.4 hours). As well as this, Mixnet also substantially
reduces the percentage of late deliveries in all cities and in some
cases almost reducing the percentage of late deliveries to nearly 0\%
such as in Boston (Figure~\ref{fig:synthetic:window:bos}). For most
cities however, we would consider an ideal amount of late deliveries
to be around 10\% like with London and Beijing. From the results we
can deduce that to achieve the ideal amount of late deliveries, we
would require at least a 200\% increase in the delivery window (6.6
hours). However for most cities, the Betweenness attack still incurs
over 60\% of late deliveries even with a 200\% increase in the
delivery window, suggesting that increasing the delivery window alone
will not resolve this attack.



\subsection{Discussion}

Driverless vehicles are expected to be foundational components of
future transport systems. Our results show that the topology of road
networks plays an important role in the security of driverless
vehicles. Thus it is important to consider network topology rather
than solely focusing on the vehicles. We have shown that any
physical-proximity attack on a driverless vehicle can be carried out
at scale, if the attacker exploits certain ``ideal'' ambush
locations. By exploiting these locations, attackers can transform
host-level attacks into a practical attack that can target one or more
fleets at the scale of an entire city.  We found such locations in
each of the twelve road networks we analysed.

We found that the mainstay of routing techniques used by driverless
vehicles -- shortest path routing (which solely focuses on efficiency)
-- is highly vulnerable to betweenness centrality attacks in all the
cities we examined. As described in Section~\ref{sec:attacks}, the
betweenness centrality of an edge is the fraction of the shortest
paths that include the edge. Thus the edges with highest betweenness
centrality are ideal ambush sites against couriers using the shortest
path routing strategy.

In contrast, mixnet routing performs significantly better than
shortest path routing. Mixnet combines the notions of routing
efficiency with randomness. Randomising a courier's route leads to
greater uncertainty on the attackers part since the courier
occasionally seeks alternatives to shortest paths to the next
destination. In comparison with shortest path routing, mixnet routing
reduces the number of edges with high betweenness centrality in the
path, thus reducing the delivery-failure. In general, we found that
delivery failure rate for mixnet was half the failure rate for
shortest path routing, in most of the cities we evaluated.

In many cities, none of the defences produced serviceable results --
even after deploying randomised defences -- a coordinated attack by
approximately 10--30 attackers, can cause between 20\% to 50\% of
deliveries to be delayed, at a minimum, considering the application of
Mixnet routing strategy. In cities like Beijing as few as 8 attackers
are able to cause significant levels of disruption. An increase in the
number of attackers reduces delivery rates approximately linearly as
the number of attackers.


Switching the routing strategy from the default (shortest path) to a
more resilient Mixnet helps, however further switching does not help
in most cases.  We found pure Nash equilibriums between attacks and
defences in London (Mixnet vs Botgrep), Beijing (Mixnet vs
Hierarchical modularity), and Boston (Mixnet vs Random). An
equilibrium predicts the strategic behaviour of attackers and
defenders, and specifically that switching from these strategy
combinations is unlikely under the assumption of rationality.

For Bristol, Birmingham, Edinburgh, Delhi, Glasgow, Chicago, and
Cambridge (UK), we found mixed-strategy equilibriums. This is due to a cyclical disruption in the dominance relationships between attack and defense strategy combinations. For instance, in the case of Edinburgh (UK), consider a courier who starts off using shortest path routing to minimise their transportation time. The attacker exploits the couriers' use of high-betweenness edges and attacks them using betweenness centrality, which is successfully countered by the courier using mixnet routing which leverages high-conductance cuts instead of shortest paths to minimise transportation time hence being robust to betweenness attack. Subsequently, the attacker switches to Eigen centrality attack, which significantly increases the \%late-deliveries under mixnet routing because of the high correlation between conductance-cut edges (used by mixnet) and high eigen-centrality edges. As a response, the courier can switch to leveraging  the shortest path routing or inverse centrality defense, to shift from using high-conductance edges. The attacker naturally switches to betweenness centrality, thus completing a cycle -- (Betweenness, Mixnet, Eigen-C, Inverse, Betweenness, $\dots$), which iterates. The cycle is stable since the dynamics of attack and defense constitutes a nash equilibrium. 

We observe from the late delivery rates in
figures~\ref{fig:synthetic:strategy},~\ref{fig:beijing:strategy}, and
~\ref{fig:london:strategy}, that in all cases, the effectiveness of
inverse centrality defense is not very different from that of the
shortest path routing, against the betweenness centrality attack. To
be clear inverse centrality is not the inverse of betweenness
centrality. It is the harmonic mean of three edge centralities
including betweenness. However, since road networks show low diversity
in degree centrality, the resulting edge choice for a route is a
function of betweenness (i.e shortest path) and eigen centrality (edge
importance as a function of other edges). Here betweenness plays the
major role as evidenced by the similar damage sustained by couriers
using either shortest-path routing or inverse-centrality routing
whilst under a betweenness centrality attack.

Many of the Nash equilibriums we observed  for the various cities we analysed are contain combinations of attacks and defenses that occur together frequently. See figures~\ref{fig:synthetic:strategy},~\ref{fig:beijing:strategy}, and
~\ref{fig:london:strategy}. For instance, a combination of  Botgrep and Betweenness attacks are found in an equilibrium with a combination of defense strategies of shortest path and mixnet routing. To understand why this occurs, we need to consider why some attacks are good at covering a broad spectrum of the attack surface. Efficient routing in so far as the strategies considered in this paper are based on two intuitions: shortest paths and high-conductance paths. Couriers using shortest path routing strategy are ambushed with high probability by attackers deployed at high-betweenness locations, and counter it by switching to mixnet routing that leverages a combination of min cuts, path randomisation, and shortest paths. However, the attacker can in turn counter that defense using botgrep which is specifically designed to uncover high-conductance cuts using a two-stage random walk method. This forces the courier to revert to the shortest-path routing, and the cycle continues.

While some defences can be effective against some attacks many challenges remain. Our work demonstrates that the well known shortest-path routing strategy will fail miserably, with delayed deliveries approaching 80--100\% in most cities we analysed, with the exception of Chicago, which has a lattice road-structure that offers slightly better resilience (60\% late deliveries). This reduction arises from the fact that outside of high betweenness centrality roads, the edges of the lattice provide a large number of alternate paths between origin-destination pairs with little diversity in their importance.

%
%
%

\subsection{Tactical aspects}

Previous work has shown that host-level attacks can be mounted via sensor saturation~\cite{masahide:jasa:1983,tseng:ijarai:2015,shin:ches:2017} or by exploiting the impact of adversarial inputs on machine-learning techniques~\cite{gardiner:2016:acmsur,tramer2018ensemble,dezfooli:2016:cvpr,weng2018evaluating,carlini:2016:ieeesp}. Up until now, no techniques have been proposed as to how attacks on individual hosts can be scaled. However, network effects of the road systems might start to change that as attackers realise they only need to be located at a fraction of possible locations to consistently ambush their targets. 

While the strategic aspects favour the attacker, do the tactical
options exist to complement it? This question can be answered by
considering whether the cost of using jamming equipment, amongst
others, in a deployed attack unit is economically viable when deployed
to disrupt a fleet of courier vehicles. From our results, we know that
an adversary only requires 25--30 mobile attack units, in order to
``cover'' an entire city.  Commercial GPS jamming equipment retails
for around \pounds 2000 and a laser gun mounted on a high-precision
industrial robot arm such as a UR3 device retails for around \pounds
9000, with cheaper alternatives available in the market. This
constitutes a burden of approximately \pounds 9000 per attack unit
excluding the mounting platform, adding up to a total budget of
roughly \pounds 250,000. The costs of moving the equipment can be
minimised by slightly increasing the total budget to statically occupy
ambush points defined by either high-betweenness or high-conductance
edges. In terms of the damage inflicted, the losses accruing from
failed promises to deliver on time has some link to repurchase
intentions. According to~\cite{chan:ssrn:2018}, the inability to
deliver on time just once can result in a reduction of 14\% of current
purchasers submitting a future order. An inability to deliver on time
twice reduces the customer base by a cumulative total of
26\%. Conservatively, assuming this results in the loss of an order of
magnitude lower loss in revenues from one city, it would mean an
indicative loss of 2-3\% of revenue per city.  Depending on the volume
of trade (given that some retailers filed tax returns of global
revenues of three figure billions of dollars) potential losses could
run into millions of pounds in the top-1000 large cities where most of
the business is done. We note however that these are very rough
calculations to examine the viability of tactical options and should
be considered no more than a sanity check. It is also worth noting that the
impact of late deliveries on customer satisfaction and future trade
depends on cultural and personal attributes, therefore generalising on
the basis of a scholarly study focused on any small part of the world
is not advisable.

\section{Related Work}
The game-theoretic background to the problem at hand lies in the search game within predator-prey games, also known as hider-seeker games. This is a zero sum game between a single predator and a single mobile prey. The predator and prey move about  in a search region. The game ends with positive payoff to the predator when it meets the prey. As a bio-inspired example, the {\em blancardella} wasp finds  larvae by searching for visible evidence of leaf-mining. Wasps are attracted by the appearance of holes or other leaf deformation created bythe leaf-mining larvae. The game begins when the wasp lands on the leaf to search for the larvae, who in turn is alerted by the vibrations caused by the landing wasp triggering evasive behaviour by larvae. When the wasp encounters a feeding hole, it repeatedly  inserts its ovipositor violently in the area to ambush the prey. The game ends either with the wasp paralysing the larvae or abandoning the leaf. The formalisation of this problem is well studied within pursuit-evasion games~\cite{adler2003randomized}.

A particular form of
hider-seeker game called an interdiction game~\cite{wood:mcm:1993}
which was originally developed to understand and intercept drug
smuggling in the 90s.  In an interdiction game, one or more smugglers
(hiders) attempt to traverse a path between two nodes on a network
while the police (seeker) patrol certain routes intensively to
interdict smugglers. Both the players are intelligent and adapt to
eachother to avoid being predictable. Our work uses Wood's game
formulation as the starting point.

Work on security games and robotic patrolling has focused on concrete
applications of {\em path-disruption}
games~\cite{bachrach2010path}. Here the hider attempts to reach a well
known target whereas the seeker wishes to prevent that. The dynamics
of attack and defence strategies is well understood in the static
target problem --- a static target is appropriately ring fenced by the
defender via a defence-in-depth approach. In our multi-party network
interdiction game, the targets are multiple as well as being dynamic
as courier deliveries involve dropoffs at numerous locations.

\section{Conclusion}

There is significant interest in using autonomous or driverless vehicles to achieve cost reductions in transport logistics of parcel deliveries and taxis to enable point-to-point transport. A major barrier to this vision is a holistic understanding of the systemic challenges across connectivity, mobility, and security. In addition to carrying out reliable data acquisition through redundant sensors, securing vehicular communications, and the host (vehicle) itself, we need carefully designed  redundancy within vehicle routing infrastructure. And, routing techniques that can leverage them via adversarially-resilient  routing algorithms. As a first step in this direction, we have carried out the first systematic analysis of the attack and defense strategy space. We showed that launching targeted attacks in an  optimal fashion is an NP-hard problem. We then applied approximation algorithms to study the dynamics of attack-defense efficiency by constructing the adversarial TSP game. We found that most of these attacks were very effective against the shortest-path routing technique which is a commonly used routing technique, while Mixnet routing was the best defense strategy. Our analysis of the adversarial TSP game identified several Nash Equilibria which offers a predictive view of which attack and defense strategies are important. While the study is not perfect in that we haven't considered the effects of congestion, we offer a lower bound of adversarial success as congestion will further reduce the fraction of on-time deliveries. There are several avenues for future work. First, our analysis would be improved by considering the effects of congestion. Second, our analysis may be improved by considering  temporal aspects (observing how variance on the traffic graphs impacts our results). Finally, we do not attempt to address the challenging problem of providing countermeasures i.e how to build redundancy into the road network and designing defensive-routing schemes that can leverage that redundancy when needed.


\bibliography{paper,biblio,botnets,unstructured-mixes}

\begin{thebibliography}{10}

\bibitem{adler2003randomized}
Micah Adler, Harald R{\"a}cke, Naveen Sivadasan, Christian Sohler, and Berthold
  V{\"o}cking.
\newblock Randomized pursuit-evasion in graphs.
\newblock {\em Combinatorics, Probability and Computing}, 12(3):225--244, 2003.

\bibitem{bachrach2010path}
Yoram Bachrach and Ely Porat.
\newblock Path disruption games.
\newblock In {\em Proceedings of the 9th International Conference on Autonomous
  Agents and Multiagent Systems: volume 1-Volume 1}, pages 1123--1130.
  International Foundation for Autonomous Agents and Multiagent Systems, 2010.

\bibitem{louvain}
Vincent~D Blondel, Jean-Loup Guillaume, Renaud Lambiotte, and Etienne Lefebvre.
\newblock Fast unfolding of communities in large networks.
\newblock {\em Journal of Statistical Mechanics: Theory and Experiment},
  (10):P10008 (12pp), 2008.

\bibitem{carlini:2016:ieeesp}
N.~{Carlini} and D.~{Wagner}.
\newblock Towards evaluating the robustness of neural networks.
\newblock {\em IEEE Transactions in Security and Privacy}, August 2017.

\bibitem{chan:ssrn:2018}
Tat Chan, Zekun Liu, and Weiqing Zhang.
\newblock Delivery service, customer satisfaction and repurchase: Evidence from
  an online retail platform.
\newblock In {\em SSRN}, 2018.

\bibitem{alg:fastgreedymodulation}
Aaron Clauset, M.~E.~J. Newman, and Cristopher Moore.
\newblock Finding community structure in very large networks.
\newblock {\em Physical Review E}, 70(6), 2004.

\bibitem{D03}
George Danezis.
\newblock Mix-networks with restricted routes.
\newblock In Roger Dingledine, editor, {\em Proceedings of Privacy Enhancing
  Technologies workshop (PET 2003)}. Springer-Verlag, LNCS 2760, March 2003.

\bibitem{ER59}
P.~Erdos and A.~Rényi.
\newblock On random graphs.
\newblock {\em Publicationes Mathemticae (Debrecen)}, 6:290--297, 1959.

\bibitem{gardiner:2016:acmsur}
Joseph Gardiner and Shishir Nagaraja.
\newblock On the security of machine learning in malware c\&\#38;c detection: A
  survey.
\newblock {\em ACM Comput. Surv.}, 49(3):59:1--59:39, December 2016.

\bibitem{dezfooli:2016:cvpr}
Seyed{-}Mohsen Moosavi{-}Dezfooli, Alhussein Fawzi, Omar Fawzi, and Pascal
  Frossard.
\newblock Universal adversarial perturbations.
\newblock {\em CoRR}, abs/1610.08401, 2016.

\bibitem{NA06}
Shishir Nagaraja and Ross Anderson.
\newblock the topology of covert conflict.
\newblock In Tyler Moore, editor, {\em Pre-Proceedings of The Fifth Workshop on
  the Economics of Information Security}, June 2006.

\bibitem{nagaraja:10:botgrep}
Shishir Nagaraja, Prateek Mittal, Chi-Yao Hong, Matthew Caesar, and Nikita
  Borisov.
\newblock {BotGrep: Finding P2P Bots with Structured Graph Analysis}.
\newblock In {\em Proc. of the USENIX Security Symposium}, 2010.

\bibitem{N06}
MEJ Newman.
\newblock {Modularity and community structure in networks}.
\newblock {\em Proceedings of the National Academy of Sciences},
  103(23):8577--8582, 2006.

\bibitem{infomap}
M.~Rosvall and C.~T. Bergstrom.
\newblock Maps of information flow reveal community structure in complex
  networks.
\newblock In {\em In Proceedings of the National Academy of Sciences USA},
  pages 1118--1123, 2007.

\bibitem{sanjab:corr:2017}
Anibal Sanjab, Walid Saad, and Tamer Basar.
\newblock Prospect theory for enhanced cyber-physical security of drone
  delivery systems: {A} network interdiction game.
\newblock {\em CoRR}, abs/1702.04240, 2017.

\bibitem{shin:ches:2017}
Hocheol Shin, Dohyun Kim, Yujin Kwon, and Yongdae Kim.
\newblock Illusion and dazzle: Adversarial optical channel exploits against
  lidars for automotive applications.
\newblock In {\em Cryptographic Hardware and Embedded Systems - {CHES} 2017 -
  19th International Conference, Taipei, Taiwan, September 25-28, 2017,
  Proceedings}, volume 10529 of {\em Lecture Notes in Computer Science}, pages
  445--467. Springer, 2017.

\bibitem{tramer2018ensemble}
Florian Tramèr, Alexey Kurakin, Nicolas Papernot, Ian Goodfellow, Dan Boneh,
  and Patrick McDaniel.
\newblock Ensemble adversarial training: Attacks and defenses.
\newblock In {\em International Conference on Learning Representations}, 2018.

\bibitem{tseng:ijarai:2015}
Wen-Kung Tseng.
\newblock A directional audible sound system using ultrasonic transducers.
\newblock 4, 09 2015.

\bibitem{weng2018evaluating}
Tsui-Wei Weng, Huan Zhang, Pin-Yu Chen, Jinfeng Yi, Dong Su, Yupeng Gao,
  Cho-Jui Hsieh, and Luca Daniel.
\newblock Evaluating the robustness of neural networks: An extreme value theory
  approach.
\newblock In {\em International Conference on Learning Representations}, 2018.

\bibitem{wood:mcm:1993}
R.Kevin Wood.
\newblock Deterministic network interdiction.
\newblock {\em Mathematical and Computer Modelling}, 17(2):1 -- 18, 1993.

\bibitem{masahide:jasa:1983}
Masahide Yoneyama, Jun‐ichiroh Fujimoto, Yu~Kawamo, and Shoichi Sasabe.
\newblock The audio spotlight: An application of nonlinear interaction of sound
  waves to a new type of loudspeaker design.
\newblock {\em The Journal of the Acoustical Society of America},
  73(5):1532--1536, 1983.

\end{thebibliography}
\bibliographystyle{plain}

\end{document}